\documentclass[11pt,a4paper]{article}
\usepackage[utf8]{inputenc}
\usepackage{textgreek}
\usepackage{jheppub}
\usepackage{enumitem}
\usepackage{booktabs}
\usepackage{microtype}
\usepackage{amsfonts}
\usepackage{mathrsfs}
\usepackage{amssymb}
\usepackage{amsmath}
\usepackage{braket}
\usepackage{hyperref} 
\usepackage{xcolor}
\usepackage{cases}
\usepackage{tikz}
\usepackage{tensor}
\usepackage{braket}
\usepackage{graphicx}
\usepackage{subcaption}

\usetikzlibrary{arrows.meta,backgrounds, decorations.pathreplacing,decorations.markings}

\newcommand{\Tr}{\text{Tr}}

\renewcommand{\i}{\text{i}}

\newcommand{\Cketbra}[1]{%
  \ket{#1}\bra{#1}_C%
}

\newcommand{\ABketbra}[1]{%
  \ket{#1}\bra{#1}_{AB}%
}

\definecolor{ink}{HTML}{263238}
\definecolor{regionA}{HTML}{DCEAF7}
\definecolor{regionB}{HTML}{F0F2F3}
\definecolor{blueA}{HTML}{2A6FBB}
\definecolor{grayB}{HTML}{90A4AE}
\definecolor{twist}{HTML}{D97721}
\definecolor{readout}{HTML}{76539B}
\definecolor{midgray}{HTML}{90A4AE}
\definecolor{softgray}{HTML}{F0F2F3}
\definecolor{sigma}{HTML}{D97721}
\definecolor{sigmaBG}{HTML}{FCE8D3}
\definecolor{even}{HTML}{2A6FBB}
\definecolor{evenBG}{HTML}{DCEAF7}
\definecolor{odd}{HTML}{76539B}
\definecolor{oddBG}{HTML}{E9E0F1}
\definecolor{cluster}{HTML}{2F8061}

\title{Detecting one-dimensional bosonic SPT phases via twisted entropic order parameter}
\author{Kosei Fujiki, Tsubasa Oishi, Soichiro Shimamori}
\affiliation{Center for Gravitational Physics and Quantum Information, Yukawa Institute for Theoretical Physics, Kyoto University,\\
	Kitashirakawa Oiwakecho, 
    Sakyo-ku, Kyoto 606-8502, Japan}
\abstract{
Entanglement asymmetry, introduced by F.~Ares, S.~Murciano and P.~Calabrese, provides a density-matrix diagnostic of symmetry breaking and successfully captures the Landau data associated with a broken symmetry pattern. However, it is by now well established that gapped quantum many-body systems can exhibit phases which are not characterized solely by Landau symmetry breaking. A fundamental example is a symmetry-protected
topological (SPT) phase, and the ordinary definition of entanglement asymmetry is insensitive to this topological information. In this work we introduce a refined quantity, which we call the twisted entropic order parameter, designed to detect SPT phases from reduced density matrices, particularly focusing on one-dimensional bosonic systems. The key ingredient in our construction is an ancilla degrees of freedom that coherently records the untwisted state and the twisted state associated to a one-ended topological defect of unbroken symmetry, so that the enlarged density matrix retains the charge carried by the defect endpoint.   
We demonstrate our proposal in concrete lattice models and further generalize it beyond ordinary group symmetries, establishing its ability to diagnose SPT phases.
This provides a first step toward a unified entanglement-asymmetry framework for diagnosing quantum phases of matter. 
}

\begin{document}
\preprint{YITP-26-95}
\maketitle
\raggedbottom

\section{Introduction}
A central organizing principle of quantum many-body physics is the Landau paradigm~\cite{Landau:1937obd}, in which phases of matter are characterized by symmetry breaking and local order parameters. In this framework, a phase is specified by a global symmetry group $G$, an unbroken subgroup $H\subset G$, and the corresponding broken-symmetry manifold $G/H$.\footnote{Throughout this paper, we assume that both $G$ and $H$ are discrete groups.} This viewpoint has been extraordinarily successful and continues to provide the basic language for describing ordered phases.

Very recently, entanglement asymmetry was introduced as a new way of diagnosing symmetry breaking from the perspective of quantum information \cite{Ares:2022koq}. To formulate this idea, one divides the total system into a subsystem $A$ and its complement $B$. For a pure state $\ket{\psi}$, the reduced density matrix on $A$ is obtained by tracing out $B$:
\begin{align}\label{eq:def_reduced}
    \rho_A
    =
    \Tr_B
    \ket{\psi}\bra{\psi}\, .
\end{align}
Given this subsystem density matrix, one compares it with the symmetrized
density matrix
\begin{align}
    \rho_{A, G}
    =
    \frac{1}{|G|}
    \sum_{g\in G}
    U_g^A
    \rho_A
    U_g^{A \dagger}\, , 
\end{align}
where $U^{A}_{g}$ is the symmetry operator defined on $A$ associated to the group element $g\in G$. The symmetrized density matrix $\rho_{A, G}$ is manifestly invariant under the $G$-action, 
\begin{align}
    U_g^A \rho_{A, G} U_g^{A\dagger} 
    =
    \rho_{A, G}\ , \quad \forall g \in G\, .  
\end{align}
The entanglement asymmetry $\Delta S_{A}$ is defined as the relative entropy between these two density matrices,
\begin{align}\label{eq:ordinary_EA}
    \Delta S_{A}
    \equiv 
    \Tr_{A}\left[\rho_{A}(\log\rho_{A} - \log\rho_{A, G})\right]\, . 
\end{align}
It therefore measures ``distance'' between the original subsystem state and the group-averaged symmetric state. It vanishes when $\rho_A$ is already invariant under $G$, while it becomes nonzero when $\rho_A$ is not invariant under $G$. For a gapped phase with spontaneous symmetry breaking $G\to H$, the large-interval behavior of $\Delta S_A$ is universal. When the size~$\ell_A$ of the
entangling region is much larger than the finite correlation length $\xi$, one finds~\cite{Capizzi:2023xaf}\footnote{
Since the symmetrized density matrix commutes with the symmetry operators, the entanglement asymmetry can be expressed as the difference between entanglement entropies associated to $\rho_A$ and $\rho_{A, G}$. Although each entropy contains an area-law UV divergence, these divergent contributions cancel each other, because the UV-divergent structure localized near the entangling surface is unchanged by the symmetrization.
}
\begin{align}\label{eq:universal_large_interval}
    \Delta S_A
    =
    \log \frac{|G|}{|H|}
    +
    O(e^{-\ell_A/\xi})\, .
\end{align}
Thus the entanglement asymmetry detects the broken orbit $G/H$. In this sense, ordinary entanglement asymmetry gives a density-matrix characterization of the Landau data of a symmetry-broken phase.\footnote{Entanglement asymmetry has also been applied to related directions such as
the quantum Mpemba effect~\cite{Ares:2023kcz, Rylands:2023yzx, Murciano:2023qrv, Yamashika:2024hpr,Chalas:2024wjz, Benini:2024xjv, Fujimura:2025rnm, Ares:2025onj},
conformal field theories~\cite{Chen:2023gql, Fossati:2024xtn, Kusuki:2024gss, Fossati:2024ekt, Lastres:2024ohf, Fossati:2026jww},
holography~\cite{Benini:2024xjv, Chen:2024lxe},
dissipative dynamics~\cite{Caceffo:2024jbc},
and generalized symmetries~\cite{Benini:2025lav, AliAhmad:2025bnd, Benini:2025hbj, Lamas:2025eay, Vescovo:2026okg, Gotta:2026tum, Zhang:2020bpf}.}

However, it has become clear over the past decade that gapped quantum many-body phases are not fully captured by the Landau paradigm alone. A particularly important class is symmetry-protected topological (SPT) phases~\cite{Gu:2009dr, Pollmann:2009ryx, Chen:2010zpc, Chen:2011pg, Chen:2011hnt, Schuch:2011niz,   Levin:2012yb, Pollmann:2012, Chen_2014}. Even when two phases have the same symmetry-breaking pattern $G\to H$, they can still be distinct if the symmetry-broken branch carries different SPT data protected by the unbroken subgroup $H$. Ordinary entanglement asymmetry defined in \eqref{eq:ordinary_EA} cannot distinguish such phases, because its large-region value~\eqref{eq:universal_large_interval} depends only on the symmetry breaking data $G/H$. This motivates the central question of this work: can one refine entanglement asymmetry so that it detects not only the broken orbit $G/H$, but also the residual SPT data protected by $H$?\footnote{
Various entanglement-based quantities have been proposed to diagnose SPT phases, including the entanglement spectrum, symmetry-resolved entanglement, charged and twisted entanglement measures, and partial-symmetry invariants~\cite{Pollmann2010EntanglementSpectrum,Li2013IdentifyingSPT,Marvian2017SPTEntanglement, Matsuura:2016qqu,Azses2020SymmetryResolved,Azses2023SymmetryResolved,Shapourian2017ManyBody,Turzillo2025Detection,Sala2026Entanglement,Sohal:2026tpv}.
Our approach is
different: twisted entropic order parameter keeps the coherent transition between
untwisted and partially twisted branches and probes the charge of a
symmetry-defect endpoint.}

As a first step toward answering this question, we restrict our attention to one-dimensional bosonic gapped phases. In this case, SPT phases protected by an onsite symmetry group $H$ are known to be classified by group cohomology $H^{2}(H, \text{U}(1))$~\cite{Chen:2011pg}. Physically, this cohomology class controls the projective quantum number carried by the endpoint of an $H$-symmetry defect.\footnote{Note that there exist nontrivial SPT phases for which no nontrivial charge is induced. Such phases cannot be detected even by string order parameters~\cite{Pollmann_2012}. In fact, when the Bogomolov multiplier \cite{Bogomolov1988, Moravec2012, Davydov:2013xov} is nontrivial, there exist SPT phases that are undetectable by either of these probes. See e.g.~\cite{Kobayashi:2025pxs, Kobayashi:2025ykb} for further details. We leave the detection of this class of SPT phases for intriguing future work.} More explicitly, if a symmetry defect labeled by $g\in H$ ends at a point, the local endpoint degree of freedom can transform projectively under the unbroken symmetry. Different cohomology classes $[\omega] \in H^{2}(H, \text{U}(1))$ therefore correspond to different ways in which symmetry-defect endpoints transform under $H$~\cite{Chen:2010zpc, Schuch:2011niz, Chen:2011hnt}. This defect-endpoint viewpoint suggests how to refine entanglement asymmetry. To access the residual $H$-SPT data, one should create a partial symmetry defect ending at the entanglement cut and ask how this endpoint transforms under the unbroken symmetry.

In this work we implement this simple idea directly at the level of reduced density matrices by introducing an ancilla degree of freedom. We introduce a refined quantity, which we call \emph{twisted entropic order parameter}. Although its precise definition and physical interpretation are discussed in the main text, the crucial point is that the ancilla qudit twisted by symmetry representation $\lambda$ allows us to extract the symmetry charge carried by a defect endpoint. For a fixed defect $g$, twisted entropic order parameter vanishes when the scanning character $\lambda$ matches the physical endpoint character $\epsilon_g^\omega$, while it is generally nonzero in the nonmatching sectors, provided that the defect coherence survives. By scanning over $\lambda$ and repeating the procedure for suitable defect labels $g$, one can extract the endpoint-charge data that distinguish the SPT phases detectable by this probe.  In this way, twisted entropic order parameter provides a density-matrix diagnostic of residual SPT order which is invisible to ordinary entanglement asymmetry.

We verify our proposal in the clock-broken $\mathbb Z_N$ cluster ladder model. In the clock-ordered regime, the symmetry is spontaneously broken as $\mathbb Z_N^{\mathrm{br}}
    \times
    \mathbb Z_N^{\mathrm e}
    \times
    \mathbb Z_N^{\mathrm o}
    \longrightarrow
    \mathbb Z_N^{\mathrm e}
    \times
    \mathbb Z_N^{\mathrm o}$
while the cluster leg realizes $N$ distinct phases, including one trivial phase and $N-1$ nontrivial SPT phases. Since all these phases share the same symmetry-breaking pattern,
ordinary entanglement asymmetry takes the same large-region value as \eqref{eq:universal_large_interval} and cannot distinguish them. By contrast, the twisted entropic order parameter resolves the symmetry charge carried by the endpoint of a partially inserted cluster
symmetry string. Scanning over the scanning characters identifies the
endpoint character and thereby distinguishes all $\mathbb Z_N$--valued SPT
indices. We demonstrate this mechanism analytically at the exactly solvable
points and perform consistency check with numerical calculations away from the
fixed-point limit.

We also explain that the same idea extends beyond ordinary group symmetries. In particular, we consider SPT phases protected by $G\times \operatorname{Rep}(G)$ symmetry~\cite{Fechisin:2023odt}. Here $\operatorname{Rep}(G)$ denotes the representation category of $G$, which becomes a genuinely categorical symmetry when $G$ is non-Abelian~\cite{Bhardwaj:2017xup, Thorngren:2019iar, Thorngren:2021yso}. We show that the twisted entropic order parameter can resolve the $G$-representation carried by the endpoint multiplet of an open $\operatorname{Rep}(G)$ symmetry operator. In contrast to the group-like case, where the endpoint response is encoded by a one-dimensional character, the endpoint multiplet may transform in a higher-dimensional irreducible representation $\Gamma$ of $G$. Our construction therefore retains the full higher-dimensional multiplet and identifies it by scanning over reference representations. This indicates that our construction is not restricted to ordinary group-SPT phases, but also applies to SPT phases protected by more general symmetry structures.

This paper is structured as follows. In section~\ref{sec:tea}, we introduce twisted entropic order parameter for one-dimensional bosonic systems with group-like symmetry-breaking patterns $G\to H$, and explain how it detects SPT data which are invisible to ordinary entanglement asymmetry. We also introduce the R\'enyi twisted entropic order parameter, which is more convenient for explicit calculations. In particular, we show that the second R\'enyi twisted entropic order parameter admits a particularly simple expression in terms of the charged and neutral components of the defect-endpoint transition operator. We also apply the general framework to the clock-broken $\mathbb Z_N$ cluster ladder model. We show that the twisted entropic order parameter distinguishes all $N$ phases. 
In section~\ref{sec:repG}, we explain how the same idea extends beyond ordinary group symmetries. We consider SPT phases protected by $G\times \operatorname{Rep}(G)$ symmetry and show that twisted entropic order parameter detects the endpoint response of open symmetry operators. This suggests that our framework can also probe SPT phases protected by onsite categorical symmetries. In section~\ref{sec:conclusion}, we summarize this paper and propose some interesting future directions. Some technical details and supplementary derivations are relegated to the appendices.

\section{Twisted entropic order parameter for group symmetries}\label{sec:tea}

In this section, we introduce the \emph{twisted entropic order parameter}, which is designed to detect the SPT data carried by a fixed symmetry-broken branch. We consider a one-dimensional bosonic gapped phase in which a finite onsite symmetry group $G$ is spontaneously broken to a subgroup $H\subset G$. After choosing a reference symmetry-broken ground state
$\ket{\psi}$ whose stabilizer is $H$, the phase is characterized not only by the broken orbit $G/H$, but also by a residual $H$-SPT class~$[\omega]\in H^2(H,\mathrm{U}(1))$.\footnote{Choosing another symmetry-broken branch,
$\ket{\psi_x}=U_x\ket{\psi}$ with $x\in G$, replaces the unbroken
subgroup by the conjugate subgroup $H_x \equiv xHx^{-1}$. The residual
SPT class is transported accordingly:
\begin{align}
    [\omega]\in H^2(H,\mathrm{U}(1))
    \quad\longrightarrow \quad
    [\omega_x]\in H^2(H_x,\mathrm{U}(1))\, , 
\end{align}
where, up to a coboundary, $\omega_x(xh_1x^{-1},xh_2x^{-1})=\omega(h_1,h_2)$.
Thus, changing the reference branch only conjugates the symmetry
labels and does not change the physical conclusions below.}
Ordinary entanglement asymmetry detects the former datum, but is insensitive to the latter in the large-region limit. The purpose of the construction below is to extract the residual SPT information from the transformation property of a symmetry-defect endpoint.

\subsection{Review of SPT phases and projective representations}
We first review how one-dimensional bosonic SPT order is encoded in the projective action of the unbroken symmetry at an entanglement cut. We recommend textbooks e.g.~\cite{Zeng:2015pxf, Tasaki:2020cpn} for more details about this topic. See also \cite{Ogata2021Classification} for rigorous discussion. To isolate a single defect endpoint, we divide an infinite chain into two complementary half-infinite regions $A$ and $B$ separated by a single entanglement cut, and choose a symmetry-broken state $\ket{\psi}_{AB}$ which remains invariant under the unbroken subgroup $H$,\footnote{In general, the symmetry-broken states are labeled by the cosets in $G/H$. Here we choose the state corresponding to the identity coset as the reference; the state labeled by $gH$ is invariant under the conjugate subgroup $gHg^{-1}$.}
\begin{align}\label{eq:U_h}
    U_h\ket{\psi}_{AB}=\ket{\psi}_{AB}\,,\quad h\in H\,.
\end{align}
Here $U_h$ is the microscopic onsite symmetry operator and forms an ordinary linear representation of~$H$. The ground state admits a Schmidt decomposition
\begin{align}\label{eq:Schmidt_decomposition}
\ket{\psi}_{AB}=\sum_{\alpha}\sqrt{\lambda_\alpha}\,\ket{\alpha}_A\ket{\alpha}_B\,,\quad
    \lambda_\alpha>0\,,\quad \sum_\alpha\lambda_\alpha=1\,,
\end{align}
so that the reduced density matrix on $A$ is $\rho_A=\sum_\alpha\lambda_\alpha\ket{\alpha}\bra{\alpha}_A$. We refer to the span of the Schmidt vectors $\{\ket{\alpha}_A\}$ as the Schmidt support of $\rho_A$.

To describe the symmetry action associated with a single cut, we use split implementers of the symmetry automorphism restricted to the two half-infinite regions, which we denote by $\widehat U_h^A$ and $\widehat U_h^B$~\cite{Schuch:2011niz, Ogata2021Classification, Else:2014vma}. They are not the bare finite tensor-product restrictions of $U_h$; a bare restriction to a finite interval remains an ordinary linear representation and carries two endpoints, whereas a split implementer carries a single endpoint at the cut.\footnote{Throughout this section, the hatted operators always denote the split half-chain implementers. In a finite-chain regularization they are obtained by terminating the remote endpoint of a long symmetry string on a boundary that explicitly breaks the relevant symmetry, see section~\ref{sec:example}.} Since only their adjoint actions are fixed, each implementer is defined up to an $h$-dependent phase. We fix the phases such that
\begin{align}\label{eq:split_global_invariance}
    \big(\widehat U_h^A\otimes\widehat U_h^B\big)\ket{\psi}_{AB}=\ket{\psi}_{AB}\,.
\end{align}
Although $U_h$ is a linear representation, the split implementers may compose only projectively,
\begin{align}\label{eq:split_projective_laws}
    \widehat U_h^A\widehat U_k^A=\omega(h,k)\,\widehat U_{hk}^A\, ,\quad
    \widehat U_h^B\widehat U_k^B=\omega(h,k)^{-1}\,\widehat U_{hk}^B\, ,\quad h,k\in H\, , 
\end{align}
with conjugate projective factors on the two sides, so that they cancel in the full symmetry action.

The invariance \eqref{eq:split_global_invariance} implies that the reduced density matrix is invariant under the split action on $A$,
\begin{align}\label{eq:invariance_UA}
    \widehat U_h^A\,\rho_A\,\widehat U_h^{A\dagger}=\rho_A\,.
\end{align}
Hence $\widehat U^A_h$ preserves the Schmidt support and acts within each degenerate eigenspace of $\rho_A$. We denote its matrix on the Schmidt support by $V_h$,
\begin{align}\label{eq:Schmidt_edge_action}
    \widehat U_h^A\ket{\alpha}_A=\sum_\beta[V_h]_{\beta\alpha}\ket{\beta}_A\,.
\end{align}
The matrices $V_h$ are unitary on the Schmidt support, commute with $\rho_A$ by \eqref{eq:invariance_UA}, and inherit the composition law of $\widehat U^A_h$,
\begin{align}\label{eq:projective_edge_representation}
    V_gV_h=\omega(g,h)\,V_{gh}\,,\quad g,h\in H\,.
\end{align}
Thus $h\mapsto V_h$ is a projective representation of $H$ on the effective degree of freedom localized at the entanglement cut. Associativity implies the two-cocycle condition
\begin{align}\label{eq:two_cocycle_condition}
    \omega(g,h)\,\omega(gh,k)=\omega(h,k)\,\omega(g,hk)\,,\quad g,h,k\in H\,.
\end{align}
The phase convention for $\widehat U^A_h$, and hence for $V_h$, is not unique: under the rephasing
\begin{align}\label{eq:edge_rephasing}
    V_h\longrightarrow V'_h=\beta(h)V_h\,,\quad \beta(h)\in\mathrm U(1)\,,
\end{align}
the cocycle changes by a coboundary,
\begin{align}\label{eq:cocycle_redefinition}
    \omega(g,h)\longrightarrow\omega'(g,h)=\frac{\beta(g)\beta(h)}{\beta(gh)}\,\omega(g,h)\,.
\end{align}
In general no choice of $\beta$ removes all projective factors, and the invariant information is the cohomology class
\begin{align}
    [\omega]\in H^2(H,\mathrm U(1))\,,
\end{align}
which labels the residual one-dimensional bosonic $H$-SPT phase~\cite{Chen:2010zpc,Schuch:2011niz,Chen:2011hnt}. In this way a single entanglement cut retains one copy of the projective representation and therefore carries the SPT information.

We now relate the projective action on the Schmidt support to the endpoint of a one-ended symmetry defect. Combining \eqref{eq:split_global_invariance} with \eqref{eq:Schmidt_edge_action}, the action of a split implementer on $B$ can be pulled through the state and traded for the edge matrix on $A$:
\begin{align}\label{eq:split_psi}
    \big(\mathbf 1_A\otimes\widehat U_g^{B\dagger}\big)\ket{\psi}_{AB}
    =\big(\widehat U_g^{A}\otimes\mathbf 1_B\big)\ket{\psi}_{AB}
    =\big(V_g\otimes\mathbf 1_B\big)\ket{\psi}_{AB}\,,
\end{align}
and taking the adjoint of the implementers,
\begin{align}\label{eq:split_psi_conjugate}
    \big(\mathbf 1_A\otimes\widehat U_g^{B}\big)\ket{\psi}_{AB}
    =\big(V_g^{\dagger}\otimes\mathbf 1_B\big)\ket{\psi}_{AB}\,.
\end{align}
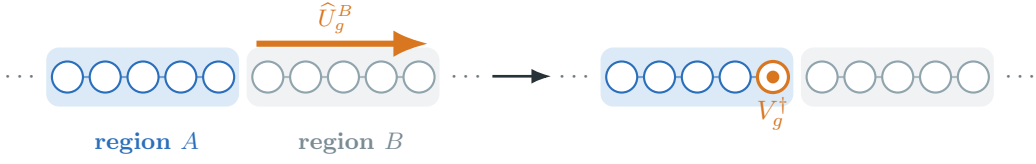
\begin{figure}
     \centering
 \begin{tikzpicture}[
     x=1cm,
     y=1cm,
     font=\small,
     text=ink,
     >={Latex[length=2.2mm,width=1.6mm]},
     siteA/.style={
       circle,
       draw=blueA,
       fill=white,
       line width=0.8pt,
       minimum size=4.1mm,
       inner sep=0pt
     },
     siteB/.style={
       circle,
       draw=grayB,
       fill=white,
       line width=0.8pt,
       minimum size=4.1mm,
       inner sep=0pt
     }
   ]
 \node[font=\footnotesize\bfseries,text=blueA] at (2.05,2.22)
   {region \(A\)};
 \node[font=\footnotesize\bfseries,text=grayB!85!black] at (4.75,2.22)
   {region \(B\)};
 \begin{scope}[on background layer]
   \fill[regionA,rounded corners=1.7mm]
     (0.72,2.70) rectangle (3.27,3.48);
   \fill[regionB,rounded corners=1.7mm]
     (3.37,2.70) rectangle (5.92,3.48);
 \end{scope}
 \draw[draw=blueA!65,line width=0.8pt] (0.96,3.09) -- (3.03,3.09);
 \draw[draw=grayB!75,line width=0.8pt] (3.61,3.09) -- (5.68,3.09);
 \foreach \x in {1.00,1.50,2.00,2.50,3.00}
   \node[siteA] at (\x,3.09) {};
\foreach \x in {3.65,4.15,4.65,5.15,5.65}
   \node[siteB] at (\x,3.09) {};
 \node[text=ink!65] at (0.37,3.09) {\(\cdots\)};
 \node[text=ink!65] at (6.27,3.09) {\(\cdots\)};
 \draw[draw=twist,line width=2.3pt,-{Latex}]
    (3.5,3.53) -- (5.78,3.53);
 \node[font=\footnotesize\bfseries,text=twist] at (4.55,3.87)
   {\(\widehat U_g^{B}\)};
 \draw[-{Latex},draw=ink,line width=1.0pt]
   (6.62,3.10) -- (7.40,3.10);
 \begin{scope}[on background layer]
   \fill[regionA,rounded corners=1.7mm]
     (8.05,2.70) rectangle (10.60,3.48);
   \fill[regionB,rounded corners=1.7mm]
     (10.70,2.70) rectangle (13.25,3.48);
 \end{scope}
 \draw[draw=blueA!65,line width=0.8pt] (8.29,3.09) -- (10.36,3.09);
 \draw[draw=grayB!75,line width=0.8pt] (10.94,3.09) -- (13.01,3.09);
 \foreach \x in {8.33,8.83,9.33,9.83}
   \node[siteA] at (\x,3.09) {};
 \node[siteA,draw=twist,line width=1.2pt] at (10.33,3.09) {};
 \foreach \x in {10.98,11.48,11.98,12.48,12.98}
   \node[siteB] at (\x,3.09) {};
 \node[text=ink!65] at (7.70,3.09) {\(\cdots\)};
 \node[text=ink!65] at (13.60,3.09) {\(\cdots\)};
 \fill[twist] (10.33,3.09) circle (2.4pt);
 \node[font=\small\bfseries,text=twist,anchor=south]
   at (10.33,2.26) {\(V_g^\dagger\)};
 \end{tikzpicture}
     \caption{Schematic illustration of a symmetry-defect endpoint charge in one-dimensional bosonic SPT fixed points. When a half-infinite $g$-symmetry operator $\widehat U_{g}^{B}$ acts on an SPT ground state, its bulk action can be pulled through the state, leaving an endpoint operator $V_g^\dagger$ localized near the right boundary of region $A$. At a fixed-point representative, this reduction is exact, while for a generic state in the same phase, the endpoint operator is dressed quasi-locally. Under the readout symmetry $\widehat U_{k}^{A}$ $(k\in C_H(g))$, its endpoint operator transforms as~\eqref{eq:endpoint_conjugation_general}.}
     \label{fig:endpoint}
 \end{figure}
\noindent Here $V_g$ is understood as the operator $\sum_{\alpha,\beta}[V_g]_{\beta\alpha}\ket{\beta}\bra{\alpha}_A$ supported on the Schmidt support. These are equalities of states: $V_g$ coincides with $\widehat U^A_g$ on the Schmidt support of $\rho_A$, but neither $V_g$ nor $V_g^\dagger$ is an operator identity with $\widehat U^{B\dagger}_g$ or $\widehat U^{B}_g$ on the microscopic Hilbert space. Physically, \eqref{eq:split_psi} states that the bulk action of the half-infinite symmetry operator $\widehat U^{B\dagger}_g$ can be pulled through the symmetric ground state, leaving the projective edge operator $V_g$ at the entanglement cut. In this sense $V_g$ is the endpoint remnant of the one-ended $g$-defect created by $\widehat U^{B\dagger}_g$, and $V_g^\dagger$ that of the defect created by $\widehat U^B_g$.\footnote{Strictly speaking, the terminology ``endpoint operator'' is accurate only at a zero-correlation-length fixed point, where $V_g$ is represented by a strictly localized operator. Away from the fixed point $V_g$ is dressed quasi-locally near the entanglement cut. We nevertheless call $V_g$ the endpoint operator throughout.} See figure~\ref{fig:endpoint}.

The endpoint charge is read out by the split implementer $\widehat U^A_k$ of an unbroken element $k\in H$. On the Schmidt support the readout acts by $V_k$, and the composition law \eqref{eq:projective_edge_representation} gives
\begin{align}\label{eq:endpoint_conjugation_general}
    \widehat U^A_k\,V_g\,\widehat U^{A\dagger}_k
    =V_kV_gV_k^{\dagger}
    =\frac{\omega(k,g)}{\omega(kgk^{-1},k)}\,V_{kgk^{-1}}\,.
\end{align}
In general the readout changes the defect species from $g$ to $kgk^{-1}$. To measure the charge of a fixed $g$-defect endpoint we therefore restrict the readout to the centralizer
\begin{align}
    C_H(g)\equiv\{k\in H\mid kgk^{-1}=g\}\,,
\end{align}
for which
\begin{align}\label{eq:endpoint_character}
    \widehat U^A_k\,V_g\,\widehat U^{A\dagger}_k=\epsilon^\omega_g(k)\,V_g\,,\quad
    \epsilon^\omega_g(k)=\frac{\omega(k,g)}{\omega(g,k)}\,,\quad k\in C_H(g)\,.
\end{align}
The phase $\epsilon^\omega_g(k)$ is the charge carried by the $g$-defect endpoint. It is a one-dimensional character of $C_H(g)$,
\begin{align}\label{eq:group}
    \epsilon^\omega_g(k)\,\epsilon^\omega_g(k')=\epsilon^\omega_g(kk')\,,\quad k,k'\in C_H(g)\,,
\end{align}
as follows from successive conjugations, and it is invariant under the rephasing \eqref{eq:edge_rephasing} and \eqref{eq:cocycle_redefinition}, since the coboundary factors $\beta(k)\beta(g)/\beta(kg)$ and $\beta(g)\beta(k)/\beta(gk)$ coincide for commuting $g,k$. Hence $\epsilon^\omega_g$ is a gauge-invariant datum determined by the class $[\omega]\in H^2(H,\mathrm U(1))$. It is the slant product of $\omega$ by $g$; appendix~\ref{sec:field_theory} for its Euclidean path-integral interpretation.

We also record the equivalent statement for the split implementers on $B$, which follows from the conjugate composition law in \eqref{eq:split_projective_laws} as an operator identity,
\begin{align}\label{eq:B_split_endpoint_character}
    \widehat U_k^{B}\,\widehat U_g^{B\dagger}\,\widehat U_k^{B\dagger}
    =\epsilon^\omega_g(k)\,\widehat U_g^{B\dagger}\,,\qquad
    \widehat U_k^{B}\,\widehat U_g^{B}\,\widehat U_k^{B\dagger}
    =\overline{\epsilon^\omega_g(k)}\,\widehat U_g^{B}\,,\qquad k\in C_H(g)\,,
\end{align}
consistently with \eqref{eq:split_psi} and \eqref{eq:split_psi_conjugate}. The twisted entropic order parameter introduced in the next subsection is designed to extract $\epsilon^\omega_g$ from the coherence between the untwisted branch $\ket{\psi}$ and the branch $\widehat U^B_g\ket{\psi}$ containing a one-ended defect.

\subsection{Twisted entropic order parameter}
We now construct an entropic quantity which probes the endpoint character $\epsilon_g^\omega$ introduced above. The essential difficulty is that the ordinary reduced density matrix does not retain the coherence between the untwisted state and the state containing a partial symmetry defect. We therefore enlarge the subsystem by introducing an ancilla qubit $\{\ket{0}_C , \ket{1}_{C}\}$ which keeps track of these two branches.
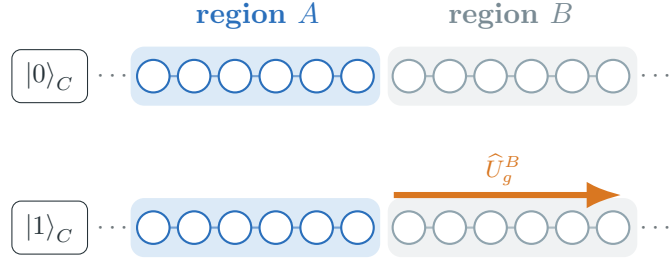
\begin{figure}[t]
\centering
\begin{tikzpicture}[
    x=1cm,
    y=1cm,
    font=\small,
    text=ink,
    >={Latex[length=2.2mm,width=1.6mm]},
    siteA/.style={
      circle,
      draw=blueA,
      fill=white,
      line width=0.8pt,
      minimum size=4.3mm,
      inner sep=0pt
    },
    siteB/.style={
      circle,
      draw=grayB,
      fill=white,
      line width=0.8pt,
      minimum size=4.3mm,
      inner sep=0pt
    },
    ancilla/.style={
      draw=ink,
      rounded corners=1.2mm,
      fill=white,
      minimum width=10mm,
      minimum height=7mm,
      inner sep=1pt
    }
  ]
\node[font=\bfseries,text=blueA] at (3.50,4.75) {region $A$};
\node[font=\bfseries,text=grayB!85!black] at (6.85,4.75) {region \(B\)};
\begin{scope}[on background layer]
  \fill[regionA,rounded corners=1.8mm]
    (1.82,3.56) rectangle (5.12,4.34);
  \fill[regionB,rounded corners=1.8mm]
    (5.22,3.56) rectangle (8.52,4.34);
  \fill[regionA,rounded corners=1.8mm]
    (1.82,1.56) rectangle (5.12,2.34);
  \fill[regionB,rounded corners=1.8mm]
    (5.22,1.56) rectangle (8.52,2.34);
\end{scope}
\node[ancilla] at (0.75,3.95) {\(\ket{0}_C\)};
\draw[draw=blueA!65,line width=0.8pt] (2.08,3.95) -- (4.88,3.95);
\draw[draw=grayB!75,line width=0.8pt] (5.46,3.95) -- (8.26,3.95);
\foreach \x in {2.12,2.66,3.20,3.74,4.28,4.82}
  \node[siteA] at (\x,3.95) {};
\foreach \x in {5.50,6.04,6.58,7.12,7.66,8.20}
  \node[siteB] at (\x,3.95) {};
\node[text=ink!65] at (1.58,3.95) {\(\cdots\)};
\node[text=ink!65] at (8.76,3.95) {\(\cdots\)};
\node[ancilla] at (0.75,1.95) {\(\ket{1}_C\)};
\draw[draw=blueA!65,line width=0.8pt] (2.08,1.95) -- (4.88,1.95);
\draw[draw=grayB!75,line width=0.8pt] (5.46,1.95) -- (8.26,1.95);
\foreach \x in {2.12,2.66,3.20,3.74,4.28,4.82}
  \node[siteA] at (\x,1.95) {};
\foreach \x in {5.50,6.04,6.58,7.12,7.66,8.20}
  \node[siteB] at (\x,1.95) {};
\node[text=ink!65] at (1.58,1.95) {\(\cdots\)};
\node[text=ink!65] at (8.76,1.95) {\(\cdots\)};
\draw[draw=twist,line width=2.3pt,-{Latex}]
   (5.30,2.38) -- (8.32,2.38);
\node[font=\footnotesize\bfseries,text=twist] at (6.75,2.72)
  {\(\widehat U_g^{B}\)};
\end{tikzpicture}
\caption{
Schematic construction of the state $\ket{\Xi_g}_{CAB}$. The ancilla state $\ket{0}_C$ labels the untwisted branch, while $\ket{1}_C$ labels the branch with the one-ended $g$-symmetry operator $\widehat U_g^{B}$ acting on region $B$.
}\label{fig:twisted_ancilla}
\end{figure}

For a fixed defect label $g\in H$, we consider the coherent
superposition
\begin{align}
\label{eq:ancilla_twisted_state}
    \ket{\Xi_g}_{CAB}
    \equiv
    \frac{1}{\sqrt{2}}
    \left(
        \ket{0}_C\ket{\psi}_{AB}
        +
        \ket{1}_C\widehat U_g^{B}\ket{\psi}_{AB}
    \right)\, .
\end{align}
Here $\ket{0}_C$ labels the untwisted branch, while $\ket{1}_C$
labels the branch containing the one-ended $g$-defect on $B$.\footnote{In a
finite-chain regularization, this object is obtained by absorbing the
remote endpoint into a boundary sector before taking the long-string limit; a bare finite
string with two endpoints is not the operator denoted by $\widehat U_g^{B}$ here.} See figure~\ref{fig:twisted_ancilla}. After
tracing out $B$, we obtain
\begin{align}
\label{eq:rho_CA_g}
    \rho_{CA}^{(g)}
    \equiv
    \operatorname{Tr}_B
    \ket{\Xi_g}\bra{\Xi_g}
    =
    \frac{1}{2}
    \begin{pmatrix}
        \rho_A & X_g \\
        X_g^\dagger & \rho_A
    \end{pmatrix}_C\, ,
\end{align}
where the block matrix is written in the ordered ancilla basis
$\{\ket{0}_C,\ket{1}_C\}$, and 
\begin{align}
\label{eq:transition_operator_Xg}
    X_g
    \equiv
    \operatorname{Tr}_B
    \left[
        \ket{\psi}\bra{\psi}_{AB}
        \widehat U_g^{B\dagger}
    \right]
\end{align}
is the transition operator between the untwisted and one-ended
$g$-defect branches. The two diagonal blocks agree because
$\widehat U_g^B$ is unitary and the partial trace is invariant under a
unitary conjugation acting entirely on $B$. The crucial point here is that the endpoint information is stored in the off-diagonal block $X_g$. The information-theoretic
construction below is meaningful for a general block $X_g$; its
identification with the universal SPT endpoint character uses the
split single-endpoint prescription specified above.

To resolve the charge content of $X_g$, we introduce a one-dimensional scanning character of the centralizer,
\begin{align}
    \lambda
    \in
    \widehat{C_H(g)}
    \equiv
    \operatorname{Hom}
    \left(
        C_H(g),
        \mathrm{U}(1)
    \right)\, .
\end{align}
We let the centralizer symmetry act also on the ancilla qubit according to
\begin{align}\label{eq:ancilla_character_action}
    W_k^{(\lambda)}
    \equiv
    \Cketbra{0}
    +
    \lambda(k)
    \Cketbra{1}\, ,
    \quad
    k\in C_H(g)\, .
\end{align}
The role of $\lambda$ will become clear in the next subsection.
The corresponding symmetry action on the enlarged Hilbert space
$\mathcal H_C\otimes\mathcal H_A$ is defined by
\begin{align}
\label{eq:enlarged_symmetry_action}
    \widetilde U_k^{(\lambda)}
    \equiv
    W_k^{(\lambda)}
    \otimes
    \widehat U_k^A\, .
\end{align}
Since $\lambda$ is a one-dimensional character whereas
$\widehat U_k^A$ is a single-edge split implementer, the enlarged
unitaries generally obey the projective multiplication law
\begin{align}\label{eq:action_U}
    \widetilde U_k^{(\lambda)}
    \widetilde U_{k'}^{(\lambda)}
    =
    \omega(k,k')
    \widetilde U_{kk'}^{(\lambda)}\, .
\end{align}
We define the character-resolved twirling channel by
\begin{align}
\label{eq:character_resolved_twirl}
    \mathcal G_{g,\lambda}^{CA}(\mathcal O)
    \equiv
    \frac{1}{|C_H(g)|}
    \sum_{k\in C_H(g)}
    \widetilde U_k^{(\lambda)}
    \mathcal O
    \widetilde U_k^{(\lambda)\dagger}\, .
\end{align}
It is a completely positive, trace-preserving, unital, and
idempotent group-twirling channel.\footnote{The projective phase in \eqref{eq:action_U} causes no difficulty for the twirling operation, because the
projective phase cancels between the unitary and its Hermitian
conjugate. Indeed, for any operator $\mathcal O$,
\begin{align}
    \widetilde U_k^{(\lambda)}
    \left(
        \widetilde U_{k'}^{(\lambda)}
        \mathcal O
        \widetilde U_{k'}^{(\lambda)\dagger}
    \right)
    \widetilde U_k^{(\lambda)\dagger}
    =
    \widetilde U_{kk'}^{(\lambda)}
    \mathcal O
    \widetilde U_{kk'}^{(\lambda)\dagger}\, .
\end{align}
} Acting with a single group element on the enlarged reduced density matrix gives
\begin{align}\label{eq:single_character_action}
    \begin{aligned}
    \widetilde U_k^{(\lambda)}
    \rho_{CA}^{(g)}
    \widetilde U_k^{(\lambda)\dagger}
    =
    \frac{1}{2}
    \begin{pmatrix}
        \rho_A
        &
        \overline{\lambda(k)}
        \widehat U_k^AX_g \widehat U_k^{A\dagger}
        \\[2mm]
        \lambda(k)
        \widehat U_k^AX_g^\dagger \widehat U_k^{A\dagger}
        &
        \rho_A
    \end{pmatrix}_C\, ,
    \end{aligned}
\end{align}
where we used \eqref{eq:invariance_UA} and the property of the one-dimensional group character. After averaging over the centralizer, we obtain
\begin{align}\label{eq:character_twirled_density_matrix}
    \mathcal G_{g,\lambda}^{CA}
    \left(
        \rho_{CA}^{(g)}
    \right)
    =
    \frac{1}{2}
    \begin{pmatrix}
        \rho_A
        &
        X_{g,\lambda}
        \\
        X_{g,\lambda}^{\dagger}
        &
        \rho_A
    \end{pmatrix}_C\, ,
\end{align}
where $X_{g,\lambda}$ is the averaged transition operator defined by
\begin{align}\label{eq:character_component_Xg}
    X_{g,\lambda}
    \equiv
    \frac{1}{|C_H(g)|}
    \sum_{k\in C_H(g)}
    \overline{\lambda(k)}
    \widehat U_k^A
    X_g
    \widehat U_k^{A\dagger}\, .
\end{align}
We then define the twisted entropic order parameter by
\begin{align}\label{eq:twisted_entropic_order_parameter}
    \Delta S_{g,\lambda}^{\mathrm{twi}}(A)
    \equiv
    D
    \left(
        \rho_{CA}^{(g)}
        \middle\|
        \mathcal G_{g,\lambda}^{CA}
        \left(
            \rho_{CA}^{(g)}
        \right)
    \right)\, ,
\end{align}
where $D(\rho\|\sigma)$ is the relative entropy between density matrices $\rho$ and $\sigma$:
\begin{align}
    D(\rho\|\sigma)
    \equiv
    \Tr
    \left[
        \rho
        \left(
            \log\rho-\log\sigma
        \right)
    \right]\, . 
\end{align}
In a similar way to the entanglement asymmetry~\cite{Ares:2022koq}, the twisted entropic order parameter~$\Delta S_{g,\lambda}^{\mathrm{twi}}(A)$ can be expressed as the difference of entanglement entropies,
\begin{align}\label{eq:twisted_EOP_EE}
    \Delta S_{g,\lambda}^{\mathrm{twi}}(A)
    =
    S_{CA}
    \left(\mathcal G_{g,\lambda}^{CA}
        \left(
            \rho_{CA}^{(g)}
        \right)
    \right)
    - 
    S_{CA}\left(\rho_{CA}^{(g)}\right)\, , 
\end{align}
where $S_{CA}(\rho)$ is the von Neumann entropy, defined by
\begin{align}
    S_{CA}(\rho)
    \equiv
    -\Tr_{CA}
    \left[\rho \log \rho \right]\, . 
\end{align}
Since the ancilla $C$ is a qubit, the twisted entropic order parameter
is bounded by
\begin{align}\label{eq:upperbound}
    0
    \leq
    \Delta S_{g,\lambda}^{\mathrm{twi}}(A)
    \leq
    \log 2\, .
\end{align}
The two limiting values have a simple physical interpretation. The
lower bound is attained when the enlarged state is invariant under the
twirl. The upper bound is attained for a
mismatched scanning character, $\lambda\neq\epsilon_g^\omega$, whenever the transition operator has the ideal single-endpoint form $X_g=\rho_AV_g$ discussed below, for
which the twirl completely removes the maximal coherence between the
untwisted and one-ended defect branches. The value of this upper bound depends on the dimension of the ancilla Hilbert space.
We relegate the proof and the
precise saturation conditions to appendix~\ref{sec:upperbound}.

\subsection{Detection of endpoint charges in SPT phases}\label{subsec:detection_spt}

We now explain why the twisted entropic order parameter detects the
charge carried by a symmetry-defect endpoint. Throughout this subsection we work in the ideal split half-chain description of section~\ref{sec:tea}, in which the identities \eqref{eq:split_psi} and \eqref{eq:split_psi_conjugate} hold exactly; at a zero-correlation-length fixed point $V_g$ is in addition a strictly localized operator.\footnote{The algebraic definition in the previous subsection applies to a general transition block
$X_g$. In a finite-chain regularization the identities below hold up to corrections that are exponentially small in the distance between the entanglement cut and the physical boundaries, see section~\ref{subsec:clock_TEOP}.}

We first evaluate the transition operator on the Schmidt support. By \eqref{eq:split_psi_conjugate}, ${}_{AB}\bra{\psi}\widehat U^{B\dagger}_g=\big(\widehat U^B_g\ket{\psi}_{AB}\big)^\dagger={}_{AB}\bra{\psi}V_g$, and since $V_g$ acts on $A$ it can be taken out of the partial trace,
\begin{align}\label{eq:Xg_fixed_point}
    X_g
    =
    \operatorname{Tr}_B
    \left[
        \ket{\psi}\bra{\psi}_{AB}V_g
    \right]
    =
    \rho_A V_g
    =
    V_g\rho_A\, ,
\end{align}
where the last equality uses $[\rho_A,V_g]=0$. Thus the transition operator is the reduced density matrix dressed by the endpoint operator, and in particular $X_g\neq0$. Combining \eqref{eq:Xg_fixed_point} with the invariance \eqref{eq:invariance_UA} and the endpoint character \eqref{eq:endpoint_character}, we find that $X_g$ transforms with the character
$\epsilon_g^\omega$ determined by the projective edge
representation,\footnote{The same result can be obtained without reference to the Schmidt support: using the pull-through identity $(\widehat U_k^A\otimes\mathbf 1_B)\ABketbra{\psi}(\widehat U_k^{A\dagger}\otimes\mathbf 1_B)=(\mathbf 1_A\otimes\widehat U_k^{B\dagger})\ABketbra{\psi}(\mathbf 1_A\otimes\widehat U_k^{B})$, which follows from \eqref{eq:split_psi}, together with the cyclicity of the partial trace over $B$, one finds
\begin{align}
\begin{aligned}
    \widehat U_k^A
    X_g
    \widehat U_k^{A\dagger}
    &=
    \operatorname{Tr}_B
    \left[
        \left(\widehat U_k^A\otimes\mathbf 1_B\right)
        \ABketbra{\psi}
        \left(\widehat U_k^{A\dagger}\otimes\mathbf 1_B\right)
        \left(\mathbf 1_A\otimes\widehat U_g^{B\dagger}\right)
    \right]
    \\
    &=\operatorname{Tr}_B
    \left[
        \ABketbra{\psi}
        \left(
            \mathbf 1_A\otimes
            \widehat U_k^B
            \widehat U_g^{B\dagger}
            \widehat U_k^{B\dagger}
        \right)
    \right]
    =
    {\epsilon_g^\omega(k)}X_g \, ,
\end{aligned}
\end{align}
where in the last step we used \eqref{eq:B_split_endpoint_character}.}
\begin{align}\label{eq:Xg_endpoint_character_general}
    \widehat U_k^A
    X_g
    \widehat U_k^{A\dagger}
    =
    \widehat U_k^A\rho_A\widehat U_k^{A\dagger}\,
    \widehat U_k^AV_g\widehat U_k^{A\dagger}
    =
    {\epsilon_g^\omega(k)}X_g \, ,
    \quad k\in C_H(g)\, .
\end{align}
Thus the transition operator carries the endpoint character
deep inside of SPT phases.

We now substitute \eqref{eq:Xg_endpoint_character_general} into the averaged transition operator \eqref{eq:character_component_Xg}. This gives
\begin{align}
    \begin{aligned}
    X_{g,\lambda}
    =
    \frac{1}{|C_H(g)|}
    \sum_{k\in C_H(g)}
    \overline{\lambda(k)}
    \epsilon_g^\omega(k)
    X_g\, .
    \end{aligned}
\end{align}
Using the orthogonality relation for one-dimensional characters,\footnote{Recall that the endpoint response $\epsilon_g^\omega$ is itself a one-dimensional unitary character of the centralizer, i.e. $\epsilon_g^\omega\in\widehat{C_H(g)}$. See the discussion around~\eqref{eq:group}. Hence the physical endpoint character and the scanning character $\lambda$ belong to the same character group.}
\begin{align}
    \frac{1}{|C_H(g)|}
    \sum_{k\in C_H(g)}
    \overline{\lambda(k)}
    \lambda'(k)
    =
    \delta_{\lambda,\lambda'}\, ,
\end{align}
we find
\begin{align}\label{eq:Xg_character_selection}
    X_{g,\lambda}
    =
    \delta_{\lambda,\epsilon_g^\omega}
    X_g\, .
\end{align}
This selection rule makes explicit the role of $\lambda$ as a scanning
character in the twisted entropic order parameter: the twirl preserves
the defect coherence $X_g$ only when $\lambda$ matches the physical
endpoint character $\epsilon_g^\omega$, while removing it in all other
character sectors. Thus scanning over
$\lambda$ identifies the endpoint charge as the unique character for
which the coherence survives. Indeed, when the scanning character agrees with the physical endpoint character, i.e. $\lambda=\epsilon_g^\omega$, we have
\begin{align}
    X_{g,\lambda}
    =
    X_g\, ,
\end{align}
and therefore
\begin{align}
    \mathcal G_{g,\epsilon_g^\omega}^{CA}
    \left(
        \rho_{CA}^{(g)}
    \right)
    =
    \rho_{CA}^{(g)}\, .
\end{align}
The faithfulness of the quantum relative entropy then implies
\begin{align}\label{eq:TEOP_matching_character_zero}
    \Delta S_{g,\epsilon_g^\omega}^{\mathrm{twi}}(A)
    =
    0\, .
\end{align}
On the other hand, for any mismatched character $\lambda\neq\epsilon_g^\omega$, the relation~\eqref{eq:Xg_character_selection} gives
\begin{align}
    X_{g,\lambda}
    =
    0\, ,
\end{align}
so that the twirled density matrix reduces to
\begin{align}
    \mathcal G_{g,\lambda}^{CA}
    \left(
        \rho_{CA}^{(g)}
    \right)
    =
    \frac{1}{2}
    \begin{pmatrix}
        \rho_A & 0
        \\
        0 & \rho_A
    \end{pmatrix}_C\, .
    \label{eq:enlargedDM_fixedpt}
\end{align}
Hence, since $X_g\neq0$,
\begin{align}
    \Delta S_{g,\lambda}^{\mathrm{twi}}(A)
    >
    0\, ,
    \qquad
    \lambda\neq\epsilon_g^\omega\, ;
\end{align}
in fact the bound \eqref{eq:upperbound} is saturated in this case, see appendix~\ref{sec:upperbound}.

In summary, by scanning the twisted entropic order parameter over all one-dimensional characters $\lambda$ of $C_H(g)$, one can identify the symmetry charge carried by the endpoint of the $g$-defect. The twisted entropic order parameter vanishes when the scanning character matches the physical endpoint character, while it is nonzero for the other character sectors, provided that the coherence between the twisted and untwisted branches survives. Repeating
this procedure for suitable defect labels $g$ determines the collection of symmetry-defect endpoint charges. Since these charges are fixed by the projective edge representation, the twisted entropic order parameter provides an entropic diagnostic of the SPT phase to which the quantum state belongs.

\subsection{R\'enyi twisted entropic order parameter}\label{subsubsec:renyitea}
Although a direct evaluation of the von Neumann entropies appearing in the twisted entropic order parameter~\eqref{eq:twisted_EOP_EE} is generally difficult, the corresponding R\'enyi entropies are often more accessible both analytically and numerically. We therefore introduce the $n$-th R\'enyi twisted entropic order parameter by
\begin{align}\label{eq:Renyi_twisted_entropic_order_parameter}
    \Delta S_{g,\lambda}^{\mathrm{twi},(n)}(A)
    \equiv
    S_{CA}^{(n)}
    \left[
        \mathcal G_{g,\lambda}^{CA}
        \left(
            \rho_{CA}^{(g)}
        \right)
    \right]
    -
    S_{CA}^{(n)}
    \left(
        \rho_{CA}^{(g)}
    \right)\, ,
\end{align}
where the $n$-th R\'enyi entropy is defined by
\begin{align}
    S_{CA}^{(n)}(\rho)
    \equiv
    \frac{1}{1-n}
    \log
    \Tr_{CA}\rho^n\, , \quad n\not = 1\, . 
\end{align}
Equivalently, the R\'enyi twisted entropic order parameter can be written as
\begin{align}\label{eq:Renyi_twisted_entropic_order_parameter_ratio}
    \Delta S_{g,\lambda}^{\mathrm{twi},(n)}(A)
    =
    \frac{1}{1-n}
    \log
    \frac{
        \Tr_{CA}
        \left[
            \left(
                \mathcal G_{g,\lambda}^{CA}
                \left(
                    \rho_{CA}^{(g)}
                \right)
            \right)^n
        \right]
    }{
        \Tr_{CA}
        \left[
            \left(
                \rho_{CA}^{(g)}
            \right)^n
        \right]
    }\, .
\end{align}
The original twisted entropic order parameter is recovered in the replica limit,
\begin{align}
    \lim_{n\to1}
    \Delta S_{g,\lambda}^{\mathrm{twi},(n)}(A)
    =
    \Delta S_{g,\lambda}^{\mathrm{twi}}(A)\, .
\end{align}
Since $\mathcal G_{g,\lambda}^{CA}$ is a mixed-unitary channel, its action makes the state more mixed in the sense of majorization. It then follows that
\begin{align}
    \Delta S_{g,\lambda}^{\mathrm{twi},(n)}(A)
    \geq 
    0\, \quad 
    \forall n>0\, .  
\end{align}
We now specialize to the second R\'enyi index, for which the result admits a particularly simple expression. We define
\begin{align}\label{eq:second_Renyi_weights}
    P
    \equiv
    \Tr_A\rho_A^2\, , \quad
    C_{\mathrm{tot}}
    \equiv
    \Tr_A
    \left[
        X_gX_g^\dagger
    \right]\, ,\quad 
    C_{\lambda}
    \equiv
    \Tr_A
    \left[
        X_{g,\lambda}
        X_{g,\lambda}^\dagger
    \right]\, .
\end{align}
Here $P$ is the purity of the ordinary reduced density matrix, $C_{\mathrm{tot}}$ measures the total coherence between the untwisted and $g$-twisted branches, and $C_{\lambda}$ is the part of this coherence carried by the character sector $\lambda$.  Using the block form of $\rho_{CA}^{(g)}$ in~\eqref{eq:rho_CA_g}, we obtain
\begin{align}\label{eq:purity_untwirled}
    \begin{aligned}
    \Tr_{CA}
    \left[
        \left(
            \rho_{CA}^{(g)}
        \right)^2
    \right]
    &=
    \frac{1}{2}
    \left(
        P+C_{\mathrm{tot}}
    \right)\, .
    \end{aligned}
\end{align}
Similarly, the character-resolved twirled density matrix~\eqref{eq:character_twirled_density_matrix} satisfies
\begin{align}\label{eq:purity_twirled}
    \Tr_{CA}
    \left[
        \left(
            \mathcal G_{g,\lambda}^{CA}
            \left(
                \rho_{CA}^{(g)}
            \right)
        \right)^2
    \right]
    =
    \frac{1}{2}
    \left(
        P+C_{\lambda}
    \right)\, .
\end{align}
Substituting these expressions into \eqref{eq:Renyi_twisted_entropic_order_parameter_ratio}, we find the exact formula
\begin{align}\label{eq:second_Renyi_TEOP_exact}
    \Delta S_{g,\lambda}^{\mathrm{twi},(2)}(A)
    =
    -\log
    \frac{
        P+C_{\lambda}
    }{
        P+C_{\mathrm{tot}}
    }\, .
\end{align}
The non-negativity of this expression is manifest. Indeed, the twirled quantity~\eqref{eq:character_component_Xg} defines the orthogonal projection of $X_g$ onto the character sector $\lambda$ with respect to the Hilbert--Schmidt inner product, which implies $0\leq C_{\lambda}\leq C_{\mathrm{tot}}$. Equality holds precisely when the transition operator lies entirely in the character sector selected by $\lambda$,
\begin{align}
    \Delta S_{g,\lambda}^{\mathrm{twi},(2)}(A)
    =
    0
    \quad\Longleftrightarrow\quad
    X_{g,\lambda}
    =
    X_g\, ,
\end{align}
as expected.

There is another useful expression of the second R\'enyi twisted entropic order parameter. By introducing
\begin{align}
    \gamma_g
    \equiv
    \frac{C_{\mathrm{tot}}}{P}\, , \quad
    w_{g,\lambda}
    \equiv
    \frac{C_{\lambda}}{C_{\mathrm{tot}}}\, ,
\end{align}
one can rewrite $\Delta S_{g,\lambda}^{\mathrm{twi},(2)}(A)$ as follows,
\begin{align}\label{eq:second_Renyi_TEOP_gamma_weight}
    \Delta S_{g,\lambda}^{\mathrm{twi},(2)}(A)
    =
    -\log
    \left(
        \frac{
            1+\gamma_g w_{g,\lambda}
        }{
            1+\gamma_g
        }
    \right)
    \, .
\end{align}
The parameter $\gamma_g$ measures the amount of surviving coherence between the two branches relative to the ordinary subsystem purity, while $w_{g,\lambda}$ is the fraction of the endpoint coherence contained in the character sector $\lambda$. In the ideal single-endpoint description of an SPT phase, the relation \eqref{eq:Xg_character_selection} gives
\begin{align}
    w_{g,\lambda}
    =
    \delta_{\lambda,\epsilon_g^\omega}\qquad 
    \mathrm{in\ SPT\ phase}\, . 
\end{align}
The second R\'enyi twisted entropic order parameter therefore reduces to
\begin{align}\label{eq:second_Renyi_TEOP_SPT}
    \Delta S_{g,\lambda}^{\mathrm{twi},(2)}(A)
    =
    \begin{cases}
        0\, ,
        &
        \lambda=\epsilon_g^\omega\, ,
        \\[2mm]
        \log(1+\gamma_g)\, ,
        &
        \lambda\neq\epsilon_g^\omega\, .
    \end{cases}
\end{align}
For the ideal split single-endpoint operator,
$X_g=\rho_AV_g$ with $V_g$ unitary and $[\rho_A,V_g]=0$, so that
$C_{\mathrm{tot}}=P$ and hence $\gamma_g=1$. In particular, this holds
at a zero-correlation-length fixed-point representative. The
nonmatching character sectors therefore take the universal value
\begin{align}
\label{eq:log2}
    \Delta S_{g,\lambda}^{\mathrm{twi},(2)}(A)
    =
    \log2\, ,
    \quad
    \lambda\neq\epsilon_g^\omega\, .
\end{align}
Values $\gamma_g<1$ may occur for a finite-distance endpoint
regularization or for a more general transition block, but should not
be attributed to the ideal split implementer itself.

Finally, the endpoint-charge diagnostic is not complete for a general
non-Abelian group~$H$. It probes the gauge-invariant commutator phases
$\epsilon_g^\omega(k)$ for commuting pairs $g,k\in H$. Cohomology
classes for which all these phases are trivial form the Bogomolov
multiplier $B_0(H)\subset H^2(H,\mathrm{U}(1))$~\cite{Bogomolov1988, Moravec2012, Davydov:2013xov}, and are therefore invisible to the twisted entropic order parameter designed in this paper.\footnote{At the level of the anomaly inflow action in $1{+}1$ dimensions, the group cohomology data~$H^2(H,\mathrm{U}(1))$ can be encoded into the slant product SPT action on the torus. Our construction of the twisted entropic order parameter captures this topological datum. On the other hand, to extract the data of the Bogomolov multiplier, one has to evaluate the SPT action on higher-genus Riemann surfaces.} Such phases can instead be detected from the full boundary projective representation, from partition functions on higher-genus surfaces, or after gauging $H$ and distinguishing the resulting $\operatorname{Rep}(H)$ symmetry-breaking phases~\cite{Kobayashi:2025pxs, Kobayashi:2025ykb, Gai:2026hjk, Warman:2026gfz}. We leave the detection of this class of SPT phases for intriguing future work. However, this obstruction is absent when $H$ is Abelian, for which $B_0(H)$ is trivial. In that case, the endpoint charges completely determine the cohomology class $[\omega]$, and our scanning procedure can detect all one-dimensional bosonic $H$-SPT phases.

\subsection{Example: clock-broken $\mathbb{Z}_N$ cluster ladder model}\label{sec:example}
In the previous subsections, we introduced the twisted entropic order parameter as a probe of the charge carried by the endpoint of a partially inserted symmetry defect. We now illustrate this mechanism in a simple family of one-dimensional lattice models. Our purpose here is not to introduce a new lattice model, but rather to provide a controlled working example in which the proposed diagnostic can be tested explicitly.

The model is designed to realize $N$ distinct phases with the same symmetry-breaking pattern but different residual SPT indices. This makes it particularly useful for our purpose: ordinary entanglement asymmetry takes the same large-region value in all these phases, while the twisted entropic order parameter can be tested against the residual SPT data in a setting where the relevant endpoint charges are analytically tractable. It would be interesting to test the twisted entropic order parameter numerically in more nontrivial interacting lattice models, where the symmetry-breaking and SPT sectors are not simply decoupled; we leave such studies for future work.

\subsubsection{Model and fixed-point endpoint charges}
\label{subsec:clock_fixed_points}
We consider a two-leg ladder in which each site carries an $N$-dimensional Hilbert space~\cite{Zhou:2003kna} with orthonormal basis $\{\ket{m}\}_{m=0}^{N-1}$ and generalized Pauli operators
\begin{align}\label{eq:clock_algebra}
    X\ket{m}=\ket{m+1}\, ,\quad Z\ket{m}=\zeta^m\ket{m}\, ,\quad \zeta\equiv e^{\frac{2\pi\i}{N}}\, ,\quad
    X^N=Z^N=\mathbf 1\, ,\quad ZX=\zeta XZ\, ,
\end{align}
where the labels are understood modulo $N$. Operators on the two legs are denoted by $X^\sigma_j,Z^\sigma_j$ and $X^\tau_j,Z^\tau_j$. The $\tau$ label is suppressed below when no confusion can arise. The Hamiltonian is $H=H_\sigma+H_\tau$ with
\begin{align}
    H_\sigma
    &=
    -J_\sigma\sum_j\left(Z_j^\sigma Z_{j+1}^{\sigma\dagger}+\mathrm{h.c.}\right)
    -h_\sigma\sum_j\left(X_j^\sigma+\mathrm{h.c.}\right)\, , 
    \label{eq:clock_sigma_H}
    \\
    H_\tau
    &=
    -\sum_{p=0}^{N-1}h_\tau^{(p)}\sum_j\left(K_{2j}^{(p)}+K_{2j+1}^{(p)}+\mathrm{h.c.}\right)\, ,
    \quad
    K_{2j}^{(p)}=Z_{2j-1}^{p}X_{2j}Z_{2j+1}^{-p}\, ,\quad
    K_{2j+1}^{(p)}=Z_{2j}^{-p}X_{2j+1}Z_{2j+2}^{p}\, ,
    \label{eq:clock_tau_H}
\end{align}
and non-negative couplings $J_\sigma,h_\sigma,h^{(p)}_\tau\geq0$. On an infinite or periodic chain, $H$ commutes with the onsite symmetry
\begin{align}\label{eq:clock_ladder_symmetry}
    G=\mathbb Z_N^{\mathrm{br}}\times\mathbb Z_N^{\mathrm e}\times\mathbb Z_N^{\mathrm o}\, ,
    \qquad
    S\equiv\prod_j X_j^\sigma\, ,\quad
    U_{\mathrm e}\equiv\prod_{j:\,\mathrm{even}}X_j^\tau\, ,\quad
    U_{\mathrm o}\equiv\prod_{j:\,\mathrm{odd}}X_j^\tau\, .
\end{align}
In the clock-ordered regime $J_\sigma\gg h_\sigma$ the $\sigma$-leg spontaneously breaks $\mathbb Z_N^{\mathrm{br}}$, and we choose the branch adiabatically connected to the product state $\ket{\Uparrow}_\sigma$ with $Z^\sigma_j=1$ on every site. Since $\mathbb Z_N^{\mathrm e}$ and $\mathbb Z_N^{\mathrm o}$ act only on the $\tau$-leg, the symmetry-breaking pattern is
\begin{align}\label{eq:clock_symmetry_breaking}
    G\longrightarrow H=\mathbb Z_N^{\mathrm e}\times\mathbb Z_N^{\mathrm o}\, ,
\end{align}
and the $\tau$-leg can realize the residual $H$-SPT phases classified by $H^2(\mathbb Z_N^{\mathrm e}\times\mathbb Z_N^{\mathrm o},\mathrm U(1))\simeq\mathbb Z_N$. For each $p\in\mathbb Z_N$ the ray $h^{(p)}_\tau>0$, $h^{(p')}_\tau=0$ $(p'\neq p)$ is a fixed-point representative: the ground state of the $\tau$-leg is the generalized cluster state
\begin{align}\label{eq:ZN_cluster_state}
    \ket{\mathrm{cluster}_p}
    =
    \prod_{j}\mathrm{CZ}_{2j-1,2j}^{(p)}\mathrm{CZ}_{2j,2j+1}^{(-p)}\ket{+}^{\otimes L}\, ,
    \quad
    \ket{+}\equiv\frac{1}{\sqrt N}\sum_{m=0}^{N-1}\ket{m}\, ,
    \quad
    \mathrm{CZ}_{j,k}^{(p)}\equiv\sum_{m=0}^{N-1}\ket{m}\bra{m}_j\otimes Z_k^{pm}\, ,
\end{align}
which satisfies $K^{(p)}_j\ket{\mathrm{cluster}_p}=\ket{\mathrm{cluster}_p}$ for all $j$. The case $p=0$ is the trivial product state and $p=1,\cdots,N-1$ are the nontrivial cluster SPT phases. Since all $N$ phases share the pattern \eqref{eq:clock_symmetry_breaking}, the ordinary entanglement asymmetry takes the same value $\Delta S_A=\log N+O(e^{-\ell_A/\xi})$ in all of them and is blind to $p$.

The index $p$ is read off from the endpoint of an open even-sublattice symmetry string. Multiplying the stabilizers $K^{(p)}_{2j}$ over a finite interval gives
\begin{align}\label{eq:ZN_open_string_identity}
    \Big(\prod_{j=\ell}^{M}X_{2j}\Big)\ket{\mathrm{cluster}_p}
    =
    Z_{2\ell-1}^{-p}Z_{2M+1}^{p}\ket{\mathrm{cluster}_p}\, ,
\end{align}
so that the string leaves $Z^{-p}$ and $Z^{p}$ at its two endpoints. As in section~\ref{sec:tea}, we take $A=\{j\le 2\ell-1\}$ and $B=\{j\ge2\ell\}$, so that the site adjacent to the cut on the $A$ side is odd, and denote by $\widehat U^B_{\mathrm e}$ the split one-ended implementer of $\mathbb Z_N^{\mathrm e}$ on $B$, obtained from \eqref{eq:ZN_open_string_identity} by absorbing the remote endpoint into the boundary sector. At the fixed point,
\begin{align}
\label{eq:ZN_half_string_identity}
    \widehat U_{\mathrm e}^{B}\ket{\mathrm{cluster}_p}
    =
    Z_{2\ell-1}^{-p}\ket{\mathrm{cluster}_p}
    =
    V_{\mathrm e}^{(p)\dagger}\ket{\mathrm{cluster}_p}\, ,
    \qquad
    V_{\mathrm e}^{(p)}\equiv Z_{2\ell-1}^{p}\, ,
\end{align}
which is the concrete realization of \eqref{eq:split_psi_conjugate}. The split implementer of $\mathbb Z_N^{\mathrm o}$ on $A$ acts on the site $2\ell-1$ by $X_{2\ell-1}$, so that, using $X^mZ^pX^{-m}=\zeta^{-mp}Z^p$,
\begin{align}\label{eq:ZN_endpoint_character}
    \big(\widehat U_{\mathrm o}^{A}\big)^m V_{\mathrm e}^{(p)}\big(\widehat U_{\mathrm o}^{A\dagger}\big)^m
    =
    \epsilon_{\mathrm{e}}^{(p)}(m)\,V_{\mathrm e}^{(p)}\, ,
    \quad
    \epsilon_{\mathrm{e}}^{(p)}(m)\equiv\zeta^{-m p}\, ,\quad p\in\mathbb Z_N\, .
\end{align}
The endpoint of the one-ended even-sublattice defect therefore carries the character $\epsilon_{\mathrm e}^{(p)}$ of $\mathbb Z_N^{\mathrm o}\subset H$.

\subsubsection{Twisted entropic order parameter and numerical results}
\label{subsec:clock_TEOP}
We now specialize section~\ref{sec:tea} to the defect $g=\mathrm e$ and the readout subgroup $K=\mathbb Z_N^{\mathrm o}$. For a ground state $\ket{\psi}_{AB}$ in the chosen branch, the enlarged state \eqref{eq:ancilla_twisted_state} and its reduced density matrix are
\begin{align}\label{eq:ZN_ancilla_state}
    \ket{\Xi_{\mathrm e}}_{CAB}
    \equiv
    \frac{1}{\sqrt2}\left(\ket{0}_C\ket{\psi}_{AB}+\ket{1}_C\widehat U_{\mathrm e}^{B}\ket{\psi}_{AB}\right)\, ,
    \quad
    \rho_{CA}^{(\mathrm e)}
    =
    \frac12
    \begin{pmatrix}
        \rho_A & X_{\mathrm e}\\
        X_{\mathrm e}^{\dagger} & \rho_A
    \end{pmatrix}_C\, ,
    \quad
    X_{\mathrm e}\equiv\Tr_B\left[\ket{\psi}\bra{\psi}_{AB}\widehat U_{\mathrm e}^{B\dagger}\right]\, ,
\end{align}
with $X_{\mathrm e}=\rho_AV^{(p)}_{\mathrm e}$ at the fixed point by \eqref{eq:ZN_half_string_identity}. The character-resolved twirl \eqref{eq:character_resolved_twirl} with scanning character $\lambda_q=\zeta^{-mq}$ is generated by 
\begin{align}
    \widetilde U^{(q)}_m=W^{(q)}_m\otimes(\widehat U^A_{\mathrm o})^m\, , \quad 
    W^{(q)}_m=\Cketbra{0}+\lambda_q(m)\Cketbra{1}\, ,
\end{align}
and replaces $X_{\mathrm e}$ by
\begin{align}\label{eq:twirling}
    X_{\mathrm{e},q}
    \equiv
    \frac{1}{N}\sum_{m=0}^{N-1}\zeta^{mq}\big(\widehat U_{\mathrm{o}}^A\big)^m X_\mathrm{e}\big(\widehat U_{\mathrm{o}}^{A\dagger}\big)^m
    =
    \delta_{q,p}\,X_{\mathrm e}
    \qquad\text{in the $\mathrm{SPT}_p$ phase}\, ,
\end{align}
where the last equality follows from \eqref{eq:ZN_endpoint_character} and the selection rule \eqref{eq:Xg_character_selection}. The resulting twisted entropic order parameter $\Delta S^{\mathrm{twi}}_{\mathrm e,q}(A)$ thus vanishes for $q=p$ and equals $\log 2$ for $q\neq p$ at the fixed point. Thus scanning $q$ over $0, \cdots,N-1$ identifies the SPT index.

\paragraph{Finite-chain regularization.}
In the numerical calculations, the half-infinite geometry is regularized by a chain of even $L$ sites with
\begin{align}
    A=\{1,\cdots,\ell_A\} \, , \quad B=\{\ell_A+1,\cdots,L\} \, , \quad \ell_A=2\ell-1\, , 
\end{align}
and the Hamiltonian
\begin{align}\label{eq:ZN_completed_open_H}
    H_\tau=-\sum_{p}h_\tau^{(p)}H_\tau^{(p)}\, ,\qquad
    H_\tau^{(p)}=X_1Z_2^p+\sum_{j=2}^{L-1}K_j^{(p)}+Z_{L-1}^pX_L+\mathrm{h.c.}\, ,
\end{align}
whose boundary terms are the stabilizers $K^{(p)}_1$ and $K^{(p)}_L$ with the missing site removed. They explicitly break $\mathbb Z_N^{\mathrm e}$ at the left edge and $\mathbb Z_N^{\mathrm o}$ at the right edge, lift the edge degeneracy, and provide the boundary sectors into which the remote endpoints of the symmetry strings are absorbed: the products
\begin{align}\label{eq:ZN_regularized_implementers}
    \widehat U_{\mathrm e}^{B}=\prod_{j\in B:\,\mathrm{even}}X_j\, ,\qquad
    \widehat U_{\mathrm o}^{A}=\prod_{j\in A:\,\mathrm{odd}}X_j
\end{align}
carry a single endpoint at the entanglement cut, and realize the split implementers of section~\ref{sec:tea} up to corrections exponentially small in $\min(\ell_A,L-\ell_A)/\xi$. At the fixed point one checks directly that $\widehat U^B_{\mathrm e}\ket{\mathrm{cluster}_p}=Z^{-p}_{2\ell-1}\ket{\mathrm{cluster}_p}$ and $\widehat U^A_{\mathrm o}\ket{\mathrm{cluster}_p}=Z^{-p}_{2\ell}\ket{\mathrm{cluster}_p}$, with no operator left at the physical boundaries. This prescription is essential: on a finite chain whose Hamiltonian commutes with the full $H$ (e.g.\ with periodic boundary conditions), the bare restrictions of the Abelian group $H$ to $A$ and $B$ commute with each other and satisfy $\widehat U^A_{\mathrm o}\ket{\psi}_{AB}=\widehat U^{B\dagger}_{\mathrm o}\ket{\psi}_{AB}$, so that $\widehat U^A_{\mathrm o}X_{\mathrm e}\widehat U^{A\dagger}_{\mathrm o}=X_{\mathrm e}$ identically and the twisted entropic order parameter would vanish for $q=0$ in every phase.

\begin{figure}[t]
    \begin{subfigure}{0.48\textwidth}
        \centering
        \includegraphics[width=\linewidth]{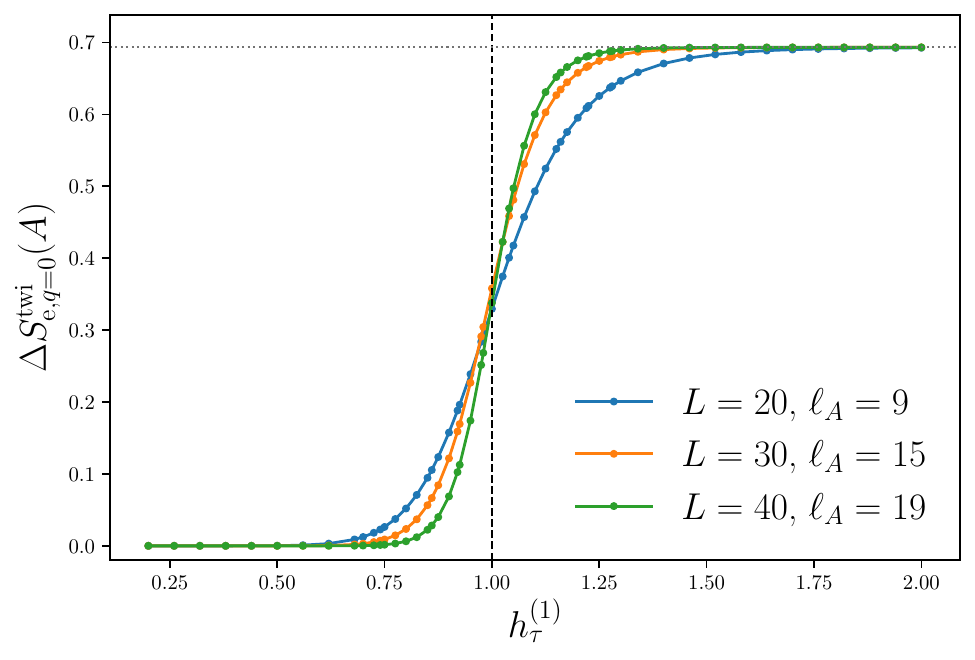}
        \caption{$q=0$}
    \end{subfigure}
    \hfill
    \begin{subfigure}{0.48\textwidth}
        \centering
        \includegraphics[width=\linewidth]{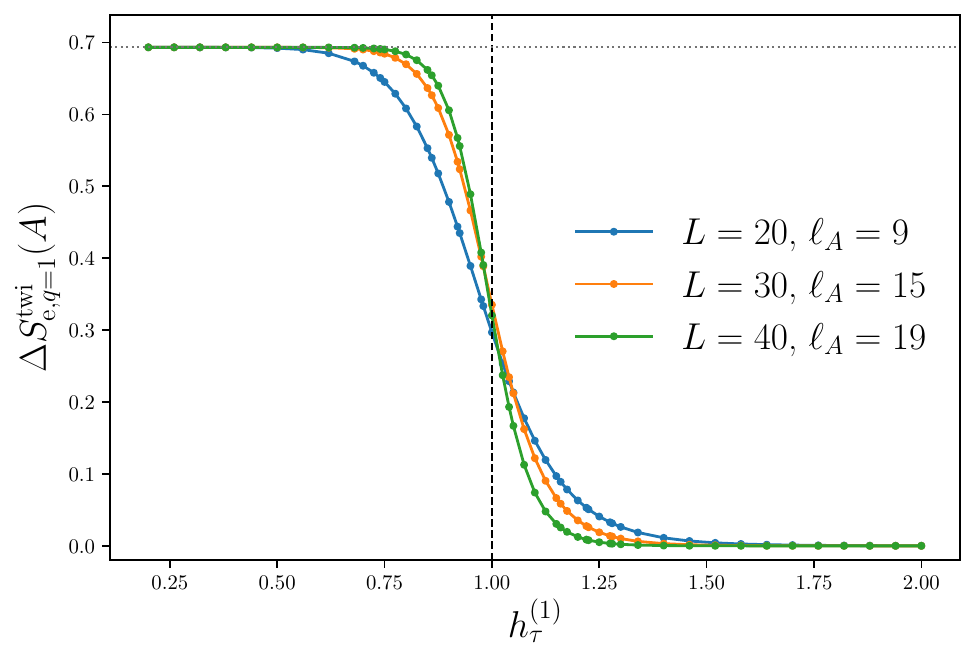}
        \caption{$q=1$}
    \end{subfigure}
    \caption{The twisted entropic order parameter for $N=2$, with $(L,\ell_A)=(20,9),(30,15),(40,19)$, computed from the ground state of \eqref{eq:ZN_completed_open_H} obtained by two-site DMRG. The transition between the trivial and SPT phases occurs at $h^{(1)}_\tau=1$.}
    \label{fig:Z2_n1_plot}
\end{figure}

\begin{figure}[t]
    \centering
    \begin{subfigure}{0.6\textwidth}
        \centering
        \includegraphics[width=\linewidth]{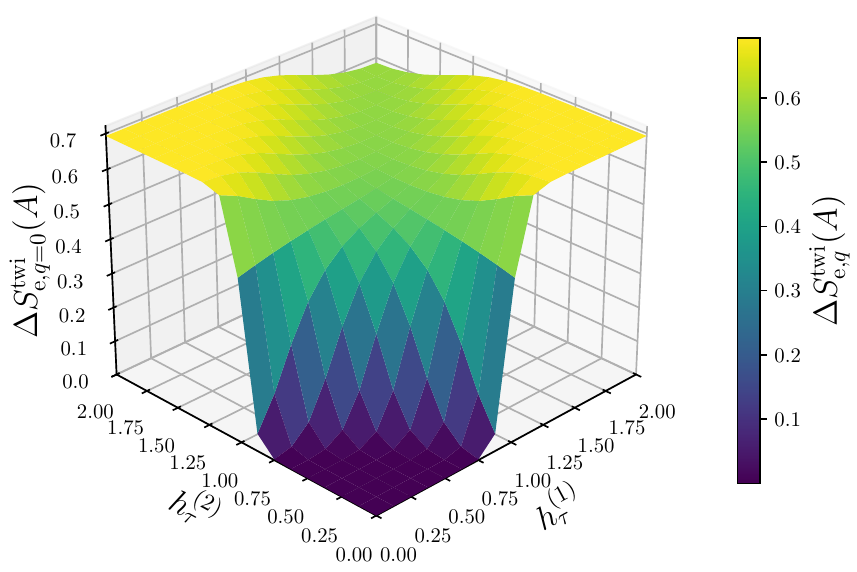}
        \caption{$q=0$}
    \end{subfigure}
    \vspace{5mm}
    \begin{subfigure}{0.48\textwidth}
        \centering
        \includegraphics[width=\linewidth]{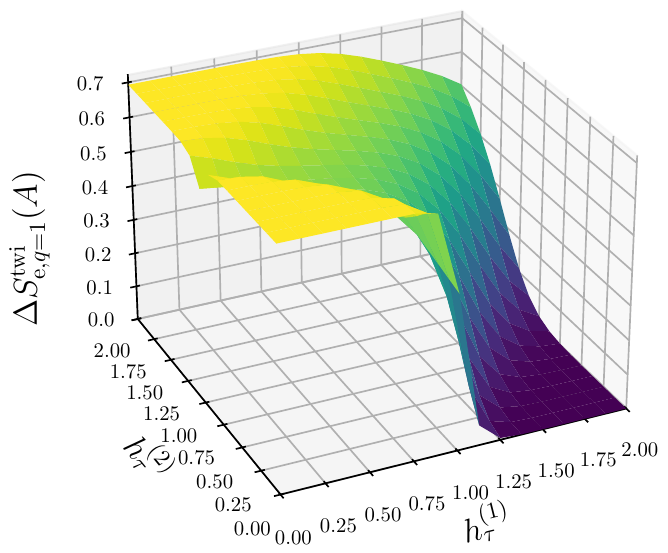}
        \caption{$q=1$}
    \end{subfigure}
    \hfill
    \begin{subfigure}{0.48\textwidth}
        \centering
        \includegraphics[width=\linewidth]{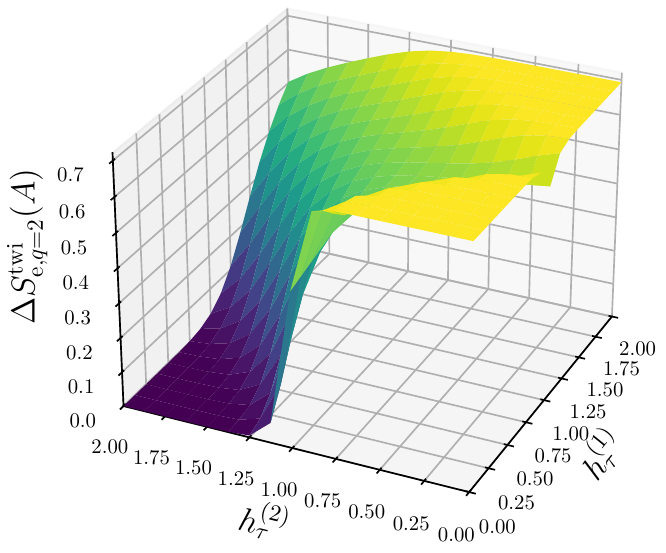}
        \caption{$q=2$}
    \end{subfigure}
    \caption{The twisted entropic order parameters for $q=0,1,2$ and $N=3$, with $L=30$ and $\ell_A=15$, computed as in figure~\ref{fig:Z2_n1_plot}. Each $\Delta S^{\mathrm{twi}}_{\mathrm e,q}$ vanishes in the $\mathrm{SPT}_q$ phase.}
    \label{fig:Z3_n1_plots}
\end{figure}

\paragraph{Numerical results.}
We set $h_{\tau}^{(0)} = 1$ without loss of generality, represent \eqref{eq:ZN_completed_open_H} as a matrix product operator, obtain the ground state by two-site DMRG, and evaluate $\Delta S^{\mathrm{twi}}_{\mathrm e,q}(A)$ with the operators \eqref{eq:ZN_regularized_implementers}. For $N=2$, figure~\ref{fig:Z2_n1_plot} shows that, across the transition at $h^{(1)}_\tau=1$, the $q=0$ order parameter changes from zero in the trivial phase to $\log2$ in the SPT phase, while the $q=1$ order parameter behaves in the opposite way, as expected. For $N=3$, figure~\ref{fig:Z3_n1_plots} shows that $\Delta S^{\mathrm{twi}}_{\mathrm e,q}$ vanishes precisely when the coupling $h^{(q)}_\tau$ dominates over the others and is nonzero elsewhere,
\begin{align}
    \Delta S_{\mathrm{e},q}^{\mathrm{twi}}\;
    \begin{cases}
         =0\, ,  & h_\tau^{(p=q)}\gg h_\tau^{(p\neq q)}\, ,\\
         \neq0 \, , & \mathrm{otherwise}\, ,
     \end{cases}
\end{align}
so that evaluating $\Delta S^{\mathrm{twi}}_{\mathrm e,q}$ for all $q$ uniquely identifies every SPT phase, in agreement with the analytical prediction.

\section{Generalized twisted entropic order parameter beyond group symmetries}\label{sec:repG}
In section~\ref{sec:tea}, we developed a general framework of twisted entropic order parameters for detecting $G$-SPT phases and illustrated it in the clock-cluster family. As both discussions were restricted to SPT phases protected by ordinary group symmetries, it is natural to ask whether the same perspective can be extended to SPT phases with more general symmetry structures, including non-invertible symmetries. In this section, we show that, at least in certain cases, this framework extends beyond ordinary group symmetries. In particular, we consider an SPT phase protected by $G\times\operatorname{Rep}(G)$ symmetry~\cite{Fechisin:2023odt} and demonstrate that its nontriviality can be detected by twisted entropic order parameters. The new ingredient is that the endpoint of an open $\operatorname{Rep}(G)$ symmetry operator carries a $G$-representation of dimension $d_\Gamma>1$ rather than a one-dimensional character. The ancilla must therefore record a whole multiplet, and the reference representation used in the twirl must be specified only up to unitary equivalence.

\subsection{$G$-based Pauli operators}
We first briefly review group-based Pauli operators acting on group-based qudits~\cite{Brell:2015vqr, Pace:2024acq, Fechisin:2023odt, Albert:2021vts, Ohyama:2026oay, Inamura:2026hjl}, mainly following~\cite{Brell:2015vqr}. A $G$-qudit is a local Hilbert space whose orthonormal basis $\ket{g}$ is labeled by the elements of a finite group $G$. The local Hilbert space is the group algebra~$\mathbb{C}[G]$, and a general state is
\begin{align}
    \ket{\psi} = \sum_{g\in G} c_g\ket{g}\, , \quad c_g\in\mathbb{C}\, .
\end{align}
For $G=\mathbb{Z}_2$, this reduces to the usual qubit. The group-based Pauli-$X$ operators act by group multiplication. Since $G$ is not assumed to be Abelian, left and right multiplications must be distinguished,
\begin{align}
    \overrightarrow{X}^{(g)} = \sum_{h\in G}\ket{gh}\bra{h}\, , \quad \overleftarrow{X}^{(g)} = \sum_{h\in G}\ket{h\bar g}\bra{h}\, , 
\end{align}
where $\bar g$ denotes the inverse of $g$. They satisfy
\begin{align}
    \overrightarrow{X}^{(g)}\overrightarrow{X}^{(h)} = \overrightarrow{X}^{(gh)}\, ,  \qquad \overleftarrow{X}^{(g)}\overleftarrow{X}^{(h)} = \overleftarrow{X}^{(gh)}\, .
\end{align}
The group-based Pauli-$Z$ operators are MPO tensors labeled by irreducible representations~$\Gamma$. Every finite-dimensional representation of a finite group is equivalent to a unitary one, and throughout this section all representations are taken unitary, $\Gamma(\bar g)=\Gamma(g)^\dagger$. We define
\begin{align}
    [\mathcal{Z}^{(\Gamma)}]_{\alpha\beta} = \sum_{g\in G}[\Gamma(g)]_{\alpha\beta}\ket{g}\bra{g}\, ,
\end{align}
where $\alpha,\beta=1,\cdots,d_\Gamma$ are indices of the $d_\Gamma$-dimensional virtual space on which $\Gamma(g)$ acts, as depicted by
\[
[\mathcal{Z}^{(\Gamma)}]_{\alpha\beta} =
\begin{tikzpicture}[baseline=-0.5ex, scale=1.0]
\node[draw, rectangle, minimum width=1.0cm, minimum height=0.8cm] (Z) at (0,0) {$\mathcal{Z}^{(\Gamma)}$};
\draw[-] (Z.north) -- (0,0.9);
\draw[-] (Z.south) -- (0,-0.9);
\draw[-, red] (Z.west) -- (-1.0,0);
\draw[-, red] (Z.east) -- (1.0,0);
\node[red] at (-1.25,0) {$\alpha$};
\node[red] at (1.25,0) {$\beta$};
\end{tikzpicture} \ , 
\]
with red lines for the virtual legs and black lines for the physical legs. An operator on the physical Hilbert space is obtained by contracting the virtual legs, e.g.\ by the trace
\begin{align}
    \text{Tr}[\mathcal{Z}^{(\Gamma)}] = \sum_{g\in G} \chi_{\Gamma}(g)\ket{g}\bra{g}\, ,
\end{align}
where $\chi_\Gamma(g)\equiv\Tr[\Gamma(g)]$ is the character. The $X$ and $Z$ operators obey
\begin{align}
    \overrightarrow{X}^{(g)}[\mathcal{Z}^{(\Gamma)}]_{\alpha\beta} &= [\Gamma(\bar g)\mathcal{Z}^{(\Gamma)}]_{\alpha\beta}\overrightarrow{X}^{(g)}\, , \\
    \overleftarrow{X}^{(g)}[\mathcal{Z}^{(\Gamma)}]_{\alpha\beta} &= [\mathcal{Z}^{(\Gamma)}\Gamma(g)]_{\alpha\beta}\overleftarrow{X}^{(g)}\, ,
\end{align}
where the representation matrices multiply the virtual indices. It is also useful to introduce the basis $\ket{\Gamma_{\alpha\beta}}$ labeled by the matrix elements of the irreducible representations,
\begin{align}
    \ket{\Gamma_{\alpha\beta}} &= \sqrt{\frac{d_{\Gamma}}{|G|}}\sum_{g\in G}[\Gamma(g)]_{\alpha\beta}\ket{g}\, ,  \\
    \ket{g}
    &=
    \sum_{\Gamma\in\operatorname{Rep}(G)}
    \sqrt{\frac{d_\Gamma}{|G|}}
    \sum_{\alpha,\beta=1}^{d_\Gamma}
    [\Gamma(\bar{g})]_{\beta\alpha}
    \ket{\Gamma_{\alpha\beta}}\, , 
\end{align}
where here and below sums over $\Gamma\in\operatorname{Rep}(G)$ run over the irreducible representations. The normalization is fixed by orthonormality,\footnote{This follows from the orthogonality relation and the character identity
\begin{align}
\frac{d_{\Gamma}}{|G|}\sum_{g\in G} \Gamma(\bar g)_{\beta\alpha}\Gamma'(g)_{\alpha'\beta'} 
= 
\delta_{\Gamma, \Gamma'}\delta_{\alpha, \alpha'}\delta_{\beta, \beta'}\, , \qquad
\frac{1}{|G|}\sum_{\Gamma \in \operatorname{Rep}(G)}d_{\Gamma} \chi_{\Gamma}(g)
=
\delta_{g, e}\, , 
\end{align}
where $e$ is the identity element of $G$.} 
\begin{align}
    \langle \Gamma_{\alpha\beta}|\Gamma'_{\gamma \varepsilon}\rangle
    =
    \delta_{\Gamma, \Gamma'}\delta_{\alpha, \gamma}\delta_{\beta, \varepsilon}\, , \quad
    \langle g | h \rangle 
    =
    \delta_{g, h}\, ,
\end{align}
and the Pauli-$X$ operators act on this basis as
\begin{align}
     \overrightarrow{X}^{(g)}\ket{\Gamma_{\alpha\beta}} = \sum_{\gamma=1}^{d_{\Gamma}}[\Gamma(\bar g)]_{\alpha\gamma}\ket{\Gamma_{\gamma\beta}}\, , \qquad
     \overleftarrow{X}^{(g)}\ket{\Gamma_{\alpha\beta}} = \sum_{\gamma=1}^{d_{\Gamma}}[\Gamma(g)]_{\gamma\beta}\ket{\Gamma_{\alpha\gamma}}\, .
\end{align}
Tracing over the virtual indices gives the character states
\begin{align}
    \ket{\Gamma} 
    \equiv
    \sum_{\alpha=1}^{d_\Gamma}
    \ket{\Gamma_{\alpha\alpha}}
    =
    \sqrt{\frac{d_{\Gamma}}{|G|}}\sum_{g\in G} \chi_{\Gamma}(g)\ket{g}\, .
\end{align}

\subsection{Generalized cluster state}\label{sec:G_cluster}
We now introduce the generalized cluster state, which extends the qubit cluster state to $G$-qudits~\cite{Brell:2015vqr, Fechisin:2023odt, Ohyama:2026oay, Inamura:2026hjl}. A further generalization to Hopf algebras is discussed in~\cite{Jia:2024bng}. We explain how it realizes an SPT phase protected by $G\times\operatorname{Rep}(G)$~\cite{Fechisin:2023odt} and, as in the group-like case, how its distinction from a product state is encoded in the symmetry charge carried by the endpoints of an open symmetry operator.\footnote{The classification of $G \times \operatorname{Rep}(G)$ SPT phases has been studied in~\cite{wang2026classificationintrinsicallymixed11d}.}

Consider a chain of $G$-qudits of even length $L$. The $G\times\operatorname{Rep}(G)$ symmetry is generated by
\begin{align} 
    G\ &:\ U_g = \prod_{j:\text{odd}}\overleftarrow{X}_{j}^{(g)}\, , \label{eq:G_sym} \\
    \operatorname{Rep}(G)\ &:\ R_{\Gamma} = \text{Tr}\big[\prod_{j:\text{even}}\mathcal{Z}_j^{(\Gamma)}\big]\, , \label{eq:Rep(G)_sym}
\end{align}
where $G$ acts on the odd sublattice and $\operatorname{Rep}(G)$ on the even sublattice. A symmetric product state is the unique ground state of
\begin{align}
    H_{\text{tri}} = -\frac{1}{|G|}\sum_{n=1}^{L/2}\Big( \sum_{\Gamma\in\operatorname{Rep}(G)}d_{\Gamma}\text{Tr}[\mathcal{Z}_{2n}^{(\Gamma)}] +\sum_{g\in G}\overrightarrow{X}_{2n+1}^{(g)} \Big)\, .
\end{align}
Using the projector identities
\begin{align}
    \frac{1}{|G|}
    \sum_{g\in G}
    \overrightarrow X^{(g)}_{2n+1}
    =
    \ket{\mathbf 1}\bra{\mathbf 1}_{2n+1}\, , \qquad
    \frac{1}{|G|}
    \sum_{\Gamma\in\operatorname{Rep}(G)}
    d_\Gamma
    \Tr\, \left[
    \mathcal Z^{(\Gamma)}_{2n}
    \right]
    =
    \ket{e}\bra{e}_{2n}\, , 
\end{align}
where $\mathbf 1$ is the trivial representation and $e$ the identity element of $G$, one has
\begin{align}
    H_{\rm tri}=-\sum_{n}\left(\ket{e}\bra{e}_{2n}+\ket{\mathbf 1}\bra{\mathbf 1}_{2n+1}\right)\, , 
\end{align} 
whose unique ground state is the product state
\begin{align}\label{eq:trivial}
    \ket{\text{Tri}} = \ket{\mathbf{1},e,\mathbf{1},e,\mathbf{1},e, \cdots}\, .
\end{align}
Since this is a product state, an open symmetry operator acting on it generates no endpoint charge. We use it as the $G\times\operatorname{Rep}(G)$-trivial reference for comparison with the generalized cluster state.\footnote{Strictly speaking, a general non-invertible SPT phase does not admit a canonical stacking operation, and hence there is no intrinsic notion of a trivial or nontrivial phase \cite{Seifnashri:2024dsd}. Here, we use the term $G\times \operatorname{Rep}(G)$-trivial phase only to mean that the corresponding open symmetry operator does not generate a nontrivial endpoint charge. Nevertheless, the absence of a canonical stacking operation does not rule out the existence of a symmetry entangler, namely, a globally symmetric finite-depth unitary circuit connecting two distinct SPT phases \cite{You:2025uxo}.}

The generalized cluster state is the unique ground state of
\begin{align}
    H_{\text{cluster}} = -\frac{1}{|G|}\sum_{n=1}^{L/2}\Big( \sum_{\Gamma\in\operatorname{Rep}(G)}d_\Gamma\,\text{Tr}[\mathcal{Z}_{2n-1}^{(\Gamma)\dagger}\mathcal{Z}_{2n}^{(\Gamma)}\mathcal{Z}_{2n+1}^{(\Gamma)}] +\sum_{g\in G}\overleftarrow{X}_{2n}^{(g)}\overrightarrow{X}_{2n+1}^{(g)}\overrightarrow{X}_{2n+2}^{(g)} \Big)\, ,
\end{align}
where $[\mathcal Z^{(\Gamma)\dagger}]_{\alpha\beta}\equiv([\mathcal Z^{(\Gamma)}]_{\beta\alpha})^\dagger=\sum_g[\Gamma(\bar g)]_{\alpha\beta}\ket g\bra g$. One checks that $H_{\rm cluster}$ commutes with \eqref{eq:G_sym} and \eqref{eq:Rep(G)_sym}, and that all its terms commute, so that the ground state is
\begin{align}\label{eq:generalized_cluster}
    \ket{\text{cluster}} = \mathcal{N}\sum_{\{g_j\}\in G}\ket{g_1, g_1\bar g_2, g_2, g_2\bar g_3,g_3, \cdots}\, ,
\end{align}
with $\mathcal N$ a normalization constant. As for the ordinary cluster state \eqref{eq:ZN_cluster_state}, it is obtained from the product state by controlled gates,
\begin{align}
    \ket{\text{cluster}} = U_{\text{cluster}}\ket{\text{Tri}}\, ,  \qquad U_{\text{cluster}} \equiv \prod_{n=1}^{L/2}\mathrm{C}\hspace{-0.5mm}\overrightarrow{\mathrm{X}}_{2n+1,2n+2}\,\mathrm{C}\hspace{-0.5mm}\overleftarrow{\mathrm{X}}_{2n+1,2n}\, ,  
\end{align}
where
\begin{align}
    \mathrm{C}\hspace{-0.5mm}\overrightarrow{\mathrm{X}}_{j,j+1}\ket{g_j, g_{j+1}} = \ket{g_j, g_jg_{j+1}}\, , \qquad
    \mathrm{C}\hspace{-0.5mm}\overleftarrow{\mathrm{X}}_{j-1,j}\ket{g_{j-1}, g_{j}} = \ket{g_{j-1}, g_{j}\bar g_{j-1}}\, .
\end{align}
The generalized cluster state satisfies the stabilizer conditions
\begin{align}\label{eq:RepG_stabilizer}
    \left[
    \mathcal{Z}_{2n-1}^{(\Gamma) \dagger}
    \mathcal{Z}_{2n}^{(\Gamma)}
    \mathcal{Z}_{2n+1}^{(\Gamma)}
    \right]_{\alpha\beta}\ket{\rm cluster}
    =
    \delta_{\alpha\beta} \ket{\rm cluster}\, , 
\end{align}
from which the action of an open $\operatorname{Rep}(G)$ operator follows immediately:
\begin{align}\label{eq:end_point_repG}
\left(R_{\Gamma}|_{\text{open}}\right)_{\alpha\beta}\ket{\text{cluster}} \equiv \left[\prod_{n=m}^{\ell}\mathcal{Z}_{2n}^{(\Gamma)}\right]_{\alpha\beta}\ket{\text{cluster}} = \big[\mathcal{Z}_{2m-1}^{(\Gamma)}\mathcal{Z}_{2\ell+1}^{(\Gamma)\dagger}\big]_{\alpha\beta}\ket{\text{cluster}}\, .
\end{align}
The open string thus leaves the operators $\mathcal Z^{(\Gamma)}_{2m-1}$ and $\mathcal Z^{(\Gamma)\dagger}_{2\ell+1}$ at its two endpoints, and these endpoint operators carry nontrivial charges under the $G$ symmetry,
\begin{align}\label{eq:RepG_endpoint_G_charge}
    U_g [\mathcal{Z}_{2m-1}^{(\Gamma)}]_{\alpha\beta}U_g^{\dagger} = [\mathcal{Z}_{2m-1}^{(\Gamma)}\Gamma(g)]_{\alpha\beta}\, ,
\end{align}
where the representation matrix acts on the second virtual index. This is the $\operatorname{Rep}(G)$ analogue of the open-string identity \eqref{eq:endpoint_conjugation_general} in section~\ref{sec:tea}, and the rest of this section shows how the charge \eqref{eq:RepG_endpoint_G_charge} is detected by a twisted entropic order parameter.

\subsection{Generalized twisted entropic order parameter}
Our strategy is the same as in the group-like case: we create a one-ended $\operatorname{Rep}(G)$ defect on $B$, record the untwisted and twisted branches in an ancilla, and read out the $G$-charge of the defect endpoint by a twirl on $A$.\footnote{One may alternatively introduce a symmetry twist on $B$ by $G$. In that case the twirling must be performed with respect to $\operatorname{Rep}(G)$, which goes beyond this paper.}

\subsubsection{One-ended $\operatorname{Rep}(G)$ defect and its endpoint multiplet}
As in section~\ref{sec:tea}, we take $A=\{j\le 2\ell-1\}$ and $B=\{j\ge 2\ell\}$, so that the site adjacent to the cut on the $A$ side, $2\ell-1$, is odd. The $\operatorname{Rep}(G)$ symmetry operator restricted to $B$ is the matrix-product operator
\begin{align}\label{eq:RepG_open_string_B}
    \big[R^{B}_\Gamma\big]_{\beta\alpha}
    \equiv
    \Big[\prod_{n\ge\ell}\mathcal Z^{(\Gamma)}_{2n}\Big]_{\beta\alpha}\, ,
\end{align}
which is a $d_\Gamma\times d_\Gamma$ matrix of operators acting on $B$. Its remote end is terminated in the same way as in section~\ref{subsec:clock_TEOP}. In a finite-chain regularization the last odd site is frozen to $\ket e$, so that the last factor $\mathcal Z^{(\Gamma)\dagger}$ in \eqref{eq:end_point_repG} reduces to the identity matrix. At the generalized-cluster fixed point the string then reduces to the endpoint operator at the cut,
\begin{align}\label{eq:RepG_string_reduction}
    \big[R^{B}_\Gamma\big]_{\beta\alpha}\ket{\mathrm{cluster}}
    =
    \big[\mathcal Z^{(\Gamma)}_{2\ell-1}\big]_{\beta\alpha}\ket{\mathrm{cluster}}\, .
\end{align}
By \eqref{eq:RepG_endpoint_G_charge} the $G$ charge acts on the second index of $\mathcal Z^{(\Gamma)}_{2\ell-1}$, which in \eqref{eq:RepG_string_reduction} is the index $\alpha$. We therefore contract the first index with a fixed boundary vector $\ket v=\sum_\beta v_\beta\ket\beta$, $\braket{v|v}=1$, and retain $\alpha$ as the label of the endpoint multiplet:
\begin{align}\label{eq:RepG_terminated_string}
    R^{B}_{\Gamma,\alpha}[v]
    \equiv
    \sum_{\beta=1}^{d_\Gamma}v_\beta\big[R^{B}_\Gamma\big]_{\beta\alpha}\, ,
    \qquad \alpha=1,\cdots,d_\Gamma\, .
\end{align}
In parallel with the group-like identity $\widehat U^B_g\ket\psi=V_g^\dagger\ket\psi$ of \eqref{eq:split_psi_conjugate}, we define the endpoint operator $V_{\Gamma,\alpha}$ by
\begin{align}\label{eq:RepG_endpoint_operator}
    V_{\Gamma,\alpha}[v]
    \equiv
    \sum_{\beta=1}^{d_\Gamma}\overline{v_\beta}\,\big[\mathcal Z^{(\Gamma)\dagger}_{2\ell-1}\big]_{\alpha\beta}
    =
    \Big(\sum_{\beta=1}^{d_\Gamma}v_\beta\big[\mathcal Z^{(\Gamma)}_{2\ell-1}\big]_{\beta\alpha}\Big)^{\dagger}\, ,
\end{align}
so that \eqref{eq:RepG_string_reduction} reads
\begin{align}\label{eq:action_R}
    R^{B}_{\Gamma,\alpha}[v]\ket{\mathrm{cluster}}
    =
    V^{\dagger}_{\Gamma,\alpha}[v]\ket{\mathrm{cluster}}\, .
\end{align}
At the product-state fixed point every site on which the string acts is in $\ket e$, and since $\Gamma(e)=\mathbf 1_{d_\Gamma}$,
\begin{align}\label{eq:R_tri}
    R^{B}_{\Gamma,\alpha}[v]\ket{\mathrm{Tri}}=v_\alpha\ket{\mathrm{Tri}}\, .
\end{align}
Taking the adjoint of \eqref{eq:RepG_endpoint_G_charge}, the endpoint multiplet transforms under $G$ as
\begin{align}\label{eq:RepG_endpoint_covariance}
    U_g\,V_{\Gamma,\alpha}\,U_g^{\dagger}
    =
    \sum_{\gamma=1}^{d_\Gamma}\big[\overline{\Gamma}(g)\big]_{\gamma\alpha}V_{\Gamma,\gamma}\, ,
    \quad
    \overline{\Gamma}(g)\equiv\overline{\Gamma(g)}\, ,
\end{align}
where $\overline\Gamma$ is the complex-conjugate representation, itself an irreducible unitary representation of $G$. We keep the boundary vector fixed and suppress the argument $[v]$ below.

\subsubsection{Enlarged density matrix}
Since the endpoint charge is now a $d_\Gamma$-dimensional representation, we retain the entire multiplet. We introduce the ancilla space
\begin{align}\label{eq:RepG_ancilla_space}
    \mathcal H^{(\Gamma)}_C
    \equiv
    \mathbb C\ket0_C\oplus\mathcal V_C\, ,
    \qquad
    \mathcal V_C\equiv\operatorname{Span}\{\ket\alpha_C\mid\alpha=1,\dots,d_\Gamma\}\, ,
\end{align}
where $\ket0_C$ labels the untwisted branch and $\ket\alpha_C$ records the component $\alpha$ of the endpoint multiplet in the twisted branch. In parallel with \eqref{eq:ancilla_twisted_state} we consider
\begin{align}\label{eq:RepG_multiplet_ancilla_state}
    \ket{\Xi_\Gamma}_{CAB}
    \equiv
    \frac{1}{\sqrt{\widetilde{\mathcal N}_\Gamma}}
    \Big[
        \ket0_C\ket\psi_{AB}
        +\sum_{\alpha=1}^{d_\Gamma}\ket\alpha_C\,R^{B}_{\Gamma,\alpha}\ket\psi_{AB}
    \Big]\, ,
    \qquad
    \widetilde{\mathcal N}_\Gamma
    \equiv
    1+\sum_{\alpha=1}^{d_\Gamma}{}_{AB}\bra{\psi} R^{B\dagger}_{\Gamma,\alpha}R^{B}_{\Gamma,\alpha}\ket{\psi}_{AB}\, .
\end{align}
Tracing out $B$ gives
\begin{align}\label{eq:RepG_multiplet_density_matrix}
    \begin{aligned}
    \rho^{(\Gamma)}_{CA}
    &\equiv\Tr_B\ket{\Xi_\Gamma}\bra{\Xi_\Gamma} \\
    &=\frac{1}{\widetilde{\mathcal N}_\Gamma}
    \Big[
        \ket0\bra0_C\otimes\rho_A
        +\sum_{\alpha}\big(\ket0\bra\alpha_C\otimes X_{\Gamma,\alpha}+\ket\alpha\bra0_C\otimes X^\dagger_{\Gamma,\alpha}\big)
        +\sum_{\alpha,\beta}\ket\alpha\bra\beta_C\otimes\rho_{A,\Gamma;\alpha\beta}
    \Big]\, ,
    \end{aligned}
\end{align}
with
\begin{align}\label{eq:RepG_multiplet_blocks}
    X_{\Gamma,\alpha}
    \equiv\Tr_B\big[\ket\psi\bra\psi_{AB}\,R^{B\dagger}_{\Gamma,\alpha}\big]\, ,
    \quad
    \rho_{A,\Gamma;\alpha\beta}
    \equiv\Tr_B\big[R^{B}_{\Gamma,\alpha}\ket\psi\bra\psi_{AB} R^{B\dagger}_{\Gamma,\beta}\big]\, .
\end{align}
These are the multiplet analogues of \eqref{eq:rho_CA_g} and \eqref{eq:transition_operator_Xg}. The transition multiplet $X_{\Gamma,\alpha}$ stores the coherence between the untwisted and twisted branches, while $\rho_{A,\Gamma;\alpha\beta}$ stores the coherences among the endpoint components within the twisted branch. At the two fixed points, \eqref{eq:action_R} and \eqref{eq:R_tri} give
\begin{align}\label{eq:RepG_transition_operator_fixed_points}
    X_{\Gamma,\alpha}=
    \begin{cases}
        \overline{v_\alpha}\,\rho_A\, , \\[1mm]
        \rho_A\,V_{\Gamma,\alpha}\, ,
    \end{cases}
    \qquad
    \rho_{A,\Gamma;\alpha\beta}=
    \begin{cases}
        v_\alpha\overline{v_\beta}\,\rho_A\, , & \text{product-state fixed point}\, ,\\[1mm]
        V^\dagger_{\Gamma,\alpha}\,\rho_A\,V_{\Gamma,\beta}\, , & \text{generalized-cluster fixed point}\, ,
    \end{cases}
\end{align}
in parallel with $X_g=\rho_AV_g$ of \eqref{eq:Xg_fixed_point}. We denote by $\Gamma_X$ the representation carried by the transition multiplet,
\begin{align}\label{eq:RepG_transition_covariance}
    U^A_g\,X_{\Gamma,\alpha}\,U^{A\dagger}_g
    =\sum_{\beta=1}^{d_\Gamma}[\Gamma_X(g)]_{\beta\alpha}X_{\Gamma,\beta}\, ,
\end{align}
where $U^A_g$ is the restriction of the onsite $G$ symmetry \eqref{eq:G_sym} to region $A$ and an ordinary linear representation.\footnote{In the group-like case both the defect and the readout belong to the same group $H$ and the endpoint charge originates from the projective composition law of the split implementers, so a bare linear restriction would miss the cocycle-dependent phase. Here the defect is an open $\operatorname{Rep}(G)$ operator and the onsite $G$ symmetry only reads out the linear $G$-representation carried by its endpoint, which is captured by the microscopic restriction $U^A_g$. On the generalized cluster state, $U^A_g\ket{\mathrm{cluster}}$ equals an operator supported on $B$ near the cut acting on $\ket{\mathrm{cluster}}$, so that $U^A_g\rho_AU^{A\dagger}_g=\rho_A$ and the conjugation in \eqref{eq:RepG_transition_covariance} is well defined; as in section~\ref{subsec:clock_TEOP}, the remote end of the readout string is absorbed at the physical boundary.} From \eqref{eq:RepG_endpoint_covariance} and \eqref{eq:RepG_transition_operator_fixed_points},
\begin{align}\label{eq:RepG_fixed_point_endpoint_representations}
    \Gamma_X=
    \begin{cases}
        \mathbf 1^{\oplus d_\Gamma}\, , & \text{product-state fixed point}\, ,\\[1mm]
        \overline\Gamma\, , & \text{generalized-cluster fixed point}\, .
    \end{cases}
\end{align}
The appearance of $\overline\Gamma$ rather than $\Gamma$ reflects the orientation of the twist. The transition block $X_{\Gamma,\alpha}$ carries the representation of the endpoint of the adjoint string $R^{B\dagger}_{\Gamma,\alpha}$, exactly as $X_g$ carries the character of the endpoint of $\widehat U^{B\dagger}_g$ in section~\ref{sec:tea}. Reversing the orientation of the twist replaces $\overline\Gamma$ by $\Gamma$.

\subsubsection{Twisted entropic order parameter for higher-dimensional endpoint charges}
To read out the endpoint representation, we let $G$ act on the charged ancilla subspace $\mathcal V_C$ through a reference representation $\Lambda$ of dimension $d_\Gamma$, in parallel with the scanning character of \eqref{eq:ancilla_character_action}. A matrix realization of $\Lambda$ is basis dependent, whereas the physically meaningful datum is its unitary-equivalence class. We therefore keep track of how the representation space $\mathcal V_\Lambda$ of $\Lambda$, with basis $\{\ket{e_a}\}_{a=1}^{d_\Gamma}$, is identified with $\mathcal V_C$. Let
\begin{align}\label{eq:RepG_identification}
    I:\ \mathcal V_\Lambda\to\mathcal V_C\, ,
    \quad
    I=\sum_{\alpha,a=1}^{d_\Gamma}I_{\alpha a}\ket\alpha_C\bra{e_a}\, ,
\end{align}
be a unitary isomorphism, and define the ancilla action
\begin{align}\label{eq:RepG_reference_action}
    W^{(\Lambda,I)}_g
    \equiv\ket0\bra0_C+\sum_{\alpha,\beta=1}^{d_\Gamma}[\Lambda_I(g)]_{\alpha\beta}\ket\alpha\bra\beta_C\, ,
    \quad
    \Lambda_I(g)\equiv I\,\Lambda(g)\,I^\dagger\, .
\end{align}
This depends only on the pair $(\Lambda,I)$ up to the simultaneous change $\Lambda\to S\Lambda S^\dagger$, $I\to IS^\dagger$ with $S\in\mathrm U(d_\Gamma)$,
\begin{align}\label{eq:RepG_I_covariance}
    W^{(S\Lambda S^\dagger,\,IS^\dagger)}_g=W^{(\Lambda,I)}_g\, ,
\end{align}
so that the family $\{W^{(\Lambda,I)}_g\}_{I\in\mathrm U(d_\Gamma)}$ depends only on the equivalence class of $\Lambda$. The action on the enlarged subsystem is
\begin{align}\label{eq:RepG_enlarged_action}
    \widetilde U^{(\Lambda,I)}_g\equiv W^{(\Lambda,I)}_g\otimes U^A_g\, ,
\end{align}
and since both factors are linear representations, $\widetilde U^{(\Lambda,I)}_g\widetilde U^{(\Lambda,I)}_h=\widetilde U^{(\Lambda,I)}_{gh}$. The representation-resolved twirling channel
\begin{align}\label{eq:RepG_representation_twirl}
    \mathcal G^{CA}_{\Gamma,\Lambda;I}(\mathcal O)
    \equiv\frac{1}{|G|}\sum_{g\in G}\widetilde U^{(\Lambda,I)}_g\,\mathcal O\,\widetilde U^{(\Lambda,I)\dagger}_g
\end{align}
is an ordinary finite-group twirl.\footnote{In general, the operation of twirling is a CPTP map. } For a fixed identification we define
\begin{align}\label{eq:RepG_TEOP_I}
    \Delta S^{\mathrm{twi}}_{\Gamma,\Lambda}(A;I)
    \equiv D\big(\rho^{(\Gamma)}_{CA}\,\big\|\,\mathcal G^{CA}_{\Gamma,\Lambda;I}(\rho^{(\Gamma)}_{CA})\big)
    =S_{CA}\big[\mathcal G^{CA}_{\Gamma,\Lambda;I}(\rho^{(\Gamma)}_{CA})\big]-S_{CA}\big(\rho^{(\Gamma)}_{CA}\big)\ \ge0\, ,
\end{align}
where the second equality holds as in \eqref{eq:twisted_EOP_EE} because the twirled state commutes with all $\widetilde U^{(\Lambda,I)}_g$. The representation-resolved twisted entropic order parameter associated with the equivalence class of $\Lambda$ is obtained by optimizing over the identification,
\begin{align}\label{eq:RepG_TEOP_definition}
    \Delta S^{\mathrm{twi}}_{\Gamma,\Lambda}(A)
    \equiv\min_{I\in\mathrm U(d_\Gamma)}\Delta S^{\mathrm{twi}}_{\Gamma,\Lambda}(A;I)\, .
\end{align}
By \eqref{eq:RepG_I_covariance} this is manifestly invariant under $\Lambda\to S\Lambda S^\dagger$, and the minimum exists because $\Delta S^{\mathrm{twi}}_{\Gamma,\Lambda}(A;I)$ is a continuous function on the compact group $\mathrm U(d_\Gamma)$. For $d_\Gamma=1$ the identification is a phase, which drops out of \eqref{eq:RepG_reference_action}, and \eqref{eq:RepG_TEOP_definition} reduces to the twisted entropic order parameter constructed in section~\ref{sec:tea}.

\subsubsection{Selection rule and detection of the endpoint representation}
Collecting the blocks into the row vector and matrix, 
\begin{align}
    \boldsymbol X_\Gamma=(X_{\Gamma,1},\dots,X_{\Gamma,d_\Gamma})\, , \quad
    [\boldsymbol\rho_{A,\Gamma}]_{\alpha\beta}=\rho_{A,\Gamma;\alpha\beta}\, , 
\end{align}
a single group element acts as
\begin{align}\label{eq:RepG_single_representation_action}
    \widetilde U^{(\Lambda,I)}_g\rho^{(\Gamma)}_{CA}\widetilde U^{(\Lambda,I)\dagger}_g
    =\frac{1}{\widetilde{\mathcal N}_\Gamma}
    \begin{pmatrix}
        \rho_A & \big(U^A_g\boldsymbol X_\Gamma U^{A\dagger}_g\big)\Lambda_I(g)^\dagger\\[1mm]
        \Lambda_I(g)\big(U^A_g\boldsymbol X^\dagger_\Gamma U^{A\dagger}_g\big) &
        \Lambda_I(g)\big(U^A_g\boldsymbol\rho_{A,\Gamma}U^{A\dagger}_g\big)\Lambda_I(g)^\dagger
    \end{pmatrix}_C\, ,
\end{align}
where $U^A_g\rho_AU^{A\dagger}_g=\rho_A$ was used. Averaging over $G$ gives rise to the following compact form, 
\begin{align}\label{eq:RepG_twirled_density_matrix}
    \mathcal G^{CA}_{\Gamma,\Lambda;I}(\rho^{(\Gamma)}_{CA})
    =\frac{1}{\widetilde{\mathcal N}_\Gamma}
    \begin{pmatrix}
        \rho_A & \boldsymbol X^{(\Lambda,I)}_\Gamma\\[1mm]
        \boldsymbol X^{(\Lambda,I)\dagger}_\Gamma & \boldsymbol\rho^{(\Lambda,I)}_{A,\Gamma}
    \end{pmatrix}_C\, ,
\end{align}
with
\begin{align}
    X^{(\Lambda,I)}_{\Gamma,\beta}
    &\equiv\frac{1}{|G|}\sum_{g\in G}\sum_{\alpha}\overline{[\Lambda_I(g)]_{\beta\alpha}}\,U^A_gX_{\Gamma,\alpha}U^{A\dagger}_g\, ,
    \label{eq:transition_multiplet}\\
    \rho^{(\Lambda,I)}_{A,\Gamma;\mu\nu}
    &\equiv\frac{1}{|G|}\sum_{g\in G}\sum_{\alpha,\beta}[\Lambda_I(g)]_{\mu\alpha}\overline{[\Lambda_I(g)]_{\nu\beta}}\,U^A_g\rho_{A,\Gamma;\alpha\beta}U^{A\dagger}_g\, .
    \label{eq:RepG_twirled_diagonal}
\end{align}
Using the covariance \eqref{eq:RepG_transition_covariance}, the twirled transition multiplet becomes
\begin{align}\label{eq:RepG_transition_projection}
    X^{(\Lambda,I)}_{\Gamma,\beta}
    =\sum_{\nu=1}^{d_\Gamma}[K_{\Gamma_X,\Lambda;I}]_{\nu\beta}X_{\Gamma,\nu}\, ,
    \quad
    K_{\Gamma_X,\Lambda;I}\equiv\frac{1}{|G|}\sum_{g\in G}\Gamma_X(g)\,\Lambda_I(g)^\dagger\, .
\end{align}
The matrix $K_{\Gamma_X,\Lambda;I}$ is an intertwiner from $\Lambda_I$ to $\Gamma_X$. Indeed, for every $h\in G$, shifting the summation variable $g\to hg$ gives
\begin{align}\label{eq:K_intertwiner}
    \Gamma_X(h)\,K_{\Gamma_X,\Lambda;I}=K_{\Gamma_X,\Lambda;I}\,\Lambda_I(h)\, .
\end{align}
Let $\Gamma_X$ be irreducible, as is the case at the fixed points \eqref{eq:RepG_fixed_point_endpoint_representations} for a nontrivial $\Gamma$. Since $\Lambda_I\simeq\Lambda$ for every $I$, Schur's lemma restricts the intertwiner \eqref{eq:K_intertwiner} as follows.
\begin{itemize}
\item If $\Lambda$ contains no irreducible component equivalent to $\Gamma_X$, i.e.\ if the multiplicity of $\Gamma_X$ in $\Lambda$ vanishes,
\begin{align}\label{eq:RepG_multiplicity}
    \langle\chi_{\Gamma_X},\chi_\Lambda\rangle_G
    \equiv\frac{1}{|G|}\sum_{g\in G}\overline{\chi_{\Gamma_X}(g)}\,\chi_\Lambda(g)=0\, ,
\end{align}
then
\begin{align}\label{eq:K_inequivalent_zero}
    K_{\Gamma_X,\Lambda;I} = 0 \qquad \forall I\, .
\end{align}
This covers an irreducible $\Lambda\not\simeq\Gamma_X$ as well as the neutral reference $\Lambda=\mathbf 1^{\oplus d_\Gamma}$, which contains only the trivial representation.
\item If $\Lambda$ is irreducible and $\Lambda\simeq\Gamma_X$, the intertwiner space is one-dimensional, and $K_{\Gamma_X,\Lambda;I}$ can be computed explicitly. For a given identification $I$ there is a unitary matrix $S\in\mathrm U(d_\Gamma)$, depending on $I$, such that $\Lambda_I(g)=S\,\Gamma_X(g)\,S^{\dagger}$. Substituting into the definition of $K_{\Gamma_X,\Lambda;I}$ in \eqref{eq:RepG_transition_projection},
\begin{align}\label{eq:RepG_K_explicit}
    K_{\Gamma_X,\Lambda;I}
    =\Big(\frac{1}{|G|}\sum_{g\in G}\Gamma_X(g)\,S\,\Gamma_X(g)^\dagger\Big)S^{\dagger}
    =\frac{\Tr S}{d_\Gamma}\,S^{\dagger}\, ,
\end{align}
where the bracket is an intertwiner of the irreducible representation $\Gamma_X$ with itself, hence proportional to $\mathbf 1_{d_\Gamma}$ by Schur's lemma, and the coefficient is fixed by taking the trace. Thus $K_{\Gamma_X,\Lambda;I}$ is proportional to the unitary matrix relating $\Lambda_I$ to $\Gamma_X$, with a coefficient $\Tr S/d_\Gamma$ that depends on the identification and vanishes for traceless $S$. Since $\Lambda$ and $\Gamma_X$ are unitarily equivalent, there exists an identification $I_\star$ for which
\begin{align}\label{eq:matched_identification}
    \Lambda_{I_\star}(g)=\Gamma_X(g)\qquad\forall g\in G\, ,
\end{align}
i.e.\ $S=\mathbf 1_{d_\Gamma}$, and for this choice \eqref{eq:RepG_K_explicit} gives $K_{\Gamma_X,\Lambda;I_\star}=\mathbf 1_{d_\Gamma}$. The identification $I_\star$ is unique up to a phase, since by Schur's lemma the only unitaries commuting with $\Gamma_X$ are multiples of the identity, and this phase drops out of \eqref{eq:RepG_reference_action}.
\end{itemize}
Using the first relation \eqref{eq:RepG_transition_projection}, we thus obtain the basis-independent selection rule for the transition multiplet,
\begin{align}\label{eq:RepG_selection_rule}
    \begin{cases}
    \boldsymbol X^{(\Lambda,I_\star)}_\Gamma=\boldsymbol X_\Gamma\quad\text{for some } I_\star\, ,
    & \Lambda\simeq\Gamma_X\, ,
    \\[2mm]
    \boldsymbol X^{(\Lambda,I)}_\Gamma=0\quad\text{for all } I\, ,
    & \langle\chi_{\Gamma_X},\chi_\Lambda\rangle_G=0\, ,
    \end{cases}
\end{align}
which is the non-Abelian analogue of the character-selection rule \eqref{eq:Xg_character_selection}. Below we simply write $\Lambda\not\simeq\Gamma_X$ for the second case $\langle\chi_{\Gamma_X},\chi_\Lambda\rangle_G=0$. The important difference is that, for a higher-dimensional representation, matching is defined only up to a unitary identification of the representation space with the charged ancilla subspace. The minimization in \eqref{eq:RepG_TEOP_definition} precisely removes this unphysical dependence.

The consequences for the twisted entropic order parameter are now immediate. At the fixed points the twisted-sector block $\boldsymbol\rho_{A,\Gamma}$ transforms with the same representation $\Gamma$.  Hence for $\Lambda\simeq\Gamma_X$ the ancilla and physical actions cancel in every block at the matched identification $I_\star$,
\begin{align}\label{eq:RepG_matched_zero}
    \mathcal G^{CA}_{\Gamma,\Lambda;I_\star}(\rho^{(\Gamma)}_{CA})=\rho^{(\Gamma)}_{CA}\, ,
    \quad
    \Delta S^{\mathrm{twi}}_{\Gamma,\Lambda}(A)=\Delta S^{\mathrm{twi}}_{\Gamma,\Lambda}(A;I_\star)=0\, ,
\end{align}
where the second equality uses the non-negativity of $\Delta S^{\mathrm{twi}}_{\Gamma,\Lambda}(A;I)$. We relegate the derivation of this equation to appendix~\ref{sec:derivation}. For $\Lambda\not\simeq\Gamma_X$, on the other hand, the transition multiplet is projected out for every $I$, so that $\mathcal G^{CA}_{\Gamma,\Lambda;I}(\rho^{(\Gamma)}_{CA})\neq\rho^{(\Gamma)}_{CA}$ whenever the endpoint coherence is nonzero, i.e. $\sum_\alpha\Tr_A[X_{\Gamma,\alpha}X^\dagger_{\Gamma,\alpha}]>0$. The faithfulness of the relative entropy then gives $\Delta S^{\mathrm{twi}}_{\Gamma,\Lambda}(A;I)>0$ for every $I$, and since a continuous positive function on the compact group $\mathrm U(d_\Gamma)$ attains its minimum,
\begin{align}\label{eq:RepG_mismatched_positive}
    \Delta S^{\mathrm{twi}}_{\Gamma,\Lambda}(A)>0\, ,
    \quad
    \Lambda\not\simeq\Gamma_X\, .
\end{align}

Combining \eqref{eq:RepG_matched_zero} and \eqref{eq:RepG_mismatched_positive} with the fixed-point values \eqref{eq:RepG_fixed_point_endpoint_representations}, we arrive at the following procedure. One evaluates $\Delta S^{\mathrm{twi}}_{\Gamma,\Lambda}(A)$ for the neutral reference $\Lambda=\mathbf 1^{\oplus d_\Gamma}$ and for each irreducible representation $\Lambda$ of dimension $d_\Gamma$. At the product-state fixed point, $\Gamma_X=\mathbf 1^{\oplus d_\Gamma}$, so the order parameter vanishes for the neutral reference and is positive for every irreducible $\Lambda$. At the generalized-cluster fixed point, $\Gamma_X=\overline\Gamma$, so the order parameter vanishes for $\Lambda\simeq\overline\Gamma$ and is positive for the neutral reference and for every other irreducible $\Lambda$. The unique $\Lambda$ for which $\Delta S^{\mathrm{twi}}_{\Gamma,\Lambda}(A)$ vanishes therefore identifies the $G$-representation carried by the endpoint of the open $\operatorname{Rep}(G)$ operator, and in particular distinguishes the generalized cluster state from the product state.\footnote{This construction identifies the endpoint $G$-representation associated with the chosen open symmetry string. Of course, we do not claim that this single-endpoint datum completely classifies all $G\times\operatorname{Rep}(G)$ SPT phases.}

\subsection{Second R\'enyi twisted entropic order parameter} 
As in section~\ref{subsubsec:renyitea}, the second R\'enyi version is the most accessible. Following \eqref{eq:Renyi_twisted_entropic_order_parameter} we define
\begin{align}\label{eq:RepG_second_Renyi_definition}
    \Delta S_{\Gamma,\Lambda}^{\mathrm{twi},(2)}(A;I)
    \equiv
    -\log
    \frac{
        \Tr_{CA}
        \big[
            \big(
                \mathcal G_{\Gamma,\Lambda;I}^{CA}
                (\rho_{CA}^{(\Gamma)})
            \big)^2
        \big]
    }{
        \Tr_{CA}
        \big[
            (\rho_{CA}^{(\Gamma)})^2
        \big]
    }\, ,
    \qquad
    \Delta S_{\Gamma,\Lambda}^{\mathrm{twi},(2)}(A)\equiv\min_{I\in\mathrm U(d_\Gamma)}\Delta S_{\Gamma,\Lambda}^{\mathrm{twi},(2)}(A;I)\, .
\end{align}
For later convenience, we introduce
\begin{align}\label{eq:RepG_second_Renyi_weights}
    \begin{aligned}
    &P
    \equiv
    \Tr_A\rho_A^2\, , \qquad
    C_{\mathrm{tot}}
    \equiv
    \sum_{\alpha=1}^{d_\Gamma}
    \Tr_A
    \big[
        X_{\Gamma,\alpha}
        X_{\Gamma,\alpha}^\dagger
    \big]\, , \qquad
    C_{\Lambda}(I)
    \equiv
    \sum_{\alpha=1}^{d_\Gamma}
    \Tr_A
    \big[
        X_{\Gamma,\alpha}^{(\Lambda,I)}
        X_{\Gamma,\alpha}^{(\Lambda,I)\dagger}
    \big]\, , \\
    &P_\Gamma
    \equiv
    \sum_{\alpha,\beta=1}^{d_\Gamma}
    \Tr_A
    \big[
        \rho_{A,\Gamma;\alpha\beta}
        \rho_{A,\Gamma;\beta\alpha}
    \big]\, , \qquad
    P_\Gamma^{(\Lambda)}(I)
    \equiv
    \sum_{\alpha,\beta=1}^{d_\Gamma}
    \Tr_A
    \big[
        \rho_{A,\Gamma;\alpha\beta}^{(\Lambda,I)}
        \rho_{A,\Gamma;\beta\alpha}^{(\Lambda,I)}
    \big]\, . 
    \end{aligned}
\end{align}
From the block forms \eqref{eq:RepG_multiplet_density_matrix} and \eqref{eq:RepG_twirled_density_matrix},
\begin{align}\label{eq:RepG_untwirled_purity}
    \Tr_{CA}\big[(\rho_{CA}^{(\Gamma)})^2\big]
    =
    \frac{P+2C_{\mathrm{tot}}+P_\Gamma}{\widetilde{\mathcal N}_\Gamma^2}\, , 
    \qquad
    \Tr_{CA}\big[\big(\mathcal G_{\Gamma,\Lambda;I}^{CA}(\rho_{CA}^{(\Gamma)})\big)^2\big]
    =
    \frac{P+2C_{\Lambda}(I)+P_\Gamma^{(\Lambda)}(I)}{\widetilde{\mathcal N}_\Gamma^2}\, ,
\end{align}
so that the normalization cancels and
\begin{align}\label{eq:RepG_second_Renyi_exact}
    \Delta S_{\Gamma,\Lambda}^{\mathrm{twi},(2)}(A;I)
    =
    -\log
    \frac{P+2C_{\Lambda}(I)+P_\Gamma^{(\Lambda)}(I)}{P+2C_{\rm tot}+P_\Gamma}\, . 
\end{align}
When $\Lambda\simeq\Gamma_X$, the matched identification $I_\star$ leaves every block invariant, $C_\Lambda(I_\star)=C_{\rm tot}$ and $P^{(\Lambda)}_\Gamma(I_\star)=P_\Gamma$, hence $\Delta S^{\mathrm{twi},(2)}_{\Gamma,\Lambda}(A)=0$. When $\Lambda\not\simeq\Gamma_X$, the selection rule \eqref{eq:RepG_selection_rule} gives $C_\Lambda(I)=0$ for every $I$, and since twirling preserves the trace of the twisted-sector block while reducing its purity  $0<P^{(\Lambda)}_\Gamma(I)\le P_\Gamma$, one finds
\begin{align}
    \Delta S_{\Gamma,\Lambda}^{\mathrm{twi},(2)}(A)
    =
    \min_{I}\left[-\log\frac{P+P_\Gamma^{(\Lambda)}(I)}{P+2C_{\rm tot}+P_\Gamma}\right]
    \ \ge\ -\log\frac{P+P_\Gamma}{P+2C_{\rm tot}+P_\Gamma}
    \ >0\, . 
\end{align}
The second R\'enyi twisted entropic order parameter therefore vanishes if and only if the reference representation matches the endpoint representation.

At the generalized-cluster fixed point with the neutral reference $\Lambda=\mathbf 1^{\oplus d_\Gamma}$, the identification drops out and the unitarity of the representation matrices gives
\begin{align}
    C_{\rm tot}=P_\Gamma=P\, , \quad
    C_{\mathbf 1^{\oplus d_\Gamma}}=0\, , \quad
    P_\Gamma^{(\mathbf 1^{\oplus d_\Gamma})}=\frac{P}{d_\Gamma}\, ,
\end{align}
and consequently
\begin{align} \label{eq:second_renyi}
    \Delta S_{\Gamma,\mathbf 1^{\oplus d_\Gamma}}^{\mathrm{twi},(2)}(A)
    =
    \log\frac{4d_\Gamma}{d_\Gamma + 1}\, . 
\end{align}
For $d_\Gamma=1$ it reduces to $\log2$, which is the group-like fixed-point value \eqref{eq:log2}.

\subsection{Example: $S_3\times \operatorname{Rep}(S_3)$ SPT phases}
We illustrate the general results with $G=S_3$, 
\begin{align}
    S_3= \langle r,s \ |\  r^3=e,\ s^2=e, \ srs = r^{-1}  \rangle
    =\{e, r,r^2, s, sr, sr^2\}\, .
\end{align}
It has three inequivalent irreducible representations, the trivial representation $\mathbf 1$, the sign representation $\mathbf 1_{\rm sign}$, and the two-dimensional representation $\mathbf 2$,
\begin{align}
    \mathbf{1}(r)=1\, , \quad \mathbf{1}(s)=1\, , \qquad
    \mathbf{1}_{\text{sign}}(r)=1\, , \quad \mathbf{1}_{\text{sign}}(s)=-1\, , \quad
    \mathbf{2}(r)=
    \begin{pmatrix}
        e^{\frac{2\pi\i}{3}} & 0 \\
        0 & e^{\frac{-2\pi\i}{3}}
    \end{pmatrix}
    , \quad \mathbf{2}(s)=
    \begin{pmatrix}
        0 & 1 \\
        1 & 0
    \end{pmatrix}
    \, ,
\end{align}
with the fusion rules
\begin{align}
\mathbf{1}_{\text{sign}}\otimes \mathbf{1}_{\text{sign}}= \mathbf{1}\, ,\quad
\mathbf{1}_{\text{sign}}\otimes \mathbf{2} = \mathbf{2}\, ,\quad
\mathbf{2}\otimes \mathbf{2}
= \mathbf{1}\oplus \mathbf{1}_{\text{sign}}\oplus \mathbf{2}\, . 
\end{align}
Note that both $\mathbf 1_{\rm sign}$ and $\mathbf 2$ are equivalent to their complex conjugates.\footnote{Explicitly, $\overline{\mathbf 2}(g)=\mathbf 2(s)\,\mathbf 2(g)\,\mathbf 2(s)$.} 

The generalized cluster state for $G=S_3$ carries two nontrivial endpoint multiplets, associated with the open $\operatorname{Rep}(S_3)$ operators $R_{\mathbf 1_{\rm sign}}$ and $R_{\mathbf 2}$, which transform under $S_3$ as in \eqref{eq:RepG_endpoint_G_charge},
\begin{align}
    U_g \mathcal{Z}_{2m-1}^{(\mathbf{1}_{\text{sign}})}U_g^{-1} = \mathcal{Z}_{2m-1}^{(\mathbf{1}_{\text{sign}})}\mathbf{1}_{\text{sign}}(g)\, , \quad 
    U_g [\mathcal{Z}_{2m-1}^{(\mathbf{2})}]_{\alpha\beta}U_g^{-1} = [\mathcal{Z}_{2m-1}^{(\mathbf{2})}\mathbf{2}(g)]_{\alpha\beta}\, .
\end{align}
By \eqref{eq:RepG_fixed_point_endpoint_representations}, the transition multiplets transform in the complex-conjugate representations, $\Gamma_X=\overline{\mathbf 1_{\rm sign}}=\mathbf 1_{\rm sign}$ for $\Gamma=\mathbf 1_{\rm sign}$ and $\Gamma_X=\overline{\mathbf 2}\simeq\mathbf 2$ for $\Gamma=\mathbf 2$. The twisted entropic order parameters therefore vanish when the reference representation is chosen accordingly,
\begin{align}
    \Delta S^{\mathrm{twi}}_{\mathbf 1_{\rm sign},\mathbf 1_{\rm sign}}(A)=0\, ,\quad
    \Delta S^{\mathrm{twi}}_{\mathbf 2,\mathbf 2}(A)=0\, .
\end{align}
The second case illustrates the role of the identification $I$. In the basis given above, $\Gamma_X=\overline{\mathbf 2}$ is related to $\Lambda=\mathbf 2$ by $\overline{\mathbf 2}(g)=\mathbf 2(s)\,\mathbf 2(g)\,\mathbf 2(s)$, so the matched identification \eqref{eq:matched_identification} is $I_\star=\mathbf 2(s)$ up to a phase, and $\Delta S^{\rm twi}_{\mathbf 2,\mathbf 2}(A;I_\star)=0$. The trivial identification $I=\mathbf 1$ behaves very differently. It corresponds to $S=\mathbf 2(s)$ in \eqref{eq:RepG_K_explicit}, and since this matrix is traceless, the intertwiner $K_{\Gamma_X,\Lambda;\mathbf 1}$ vanishes. The transition multiplet is then completely projected out by the twirl, and $\Delta S^{\rm twi}_{\mathbf 2,\mathbf 2}(A;\mathbf 1)$ is strictly positive although $\Lambda\simeq\Gamma_X$. Only after minimizing over $I$ does the order parameter correctly vanish.

For the neutral references, \eqref{eq:second_renyi} gives
\begin{align}
   \Delta S_{\mathbf{1}_{\text{sign}},\mathbf 1}^{\mathrm{twi},(2)}(A)
    =
    \log2\, ,\quad 
    \Delta S_{\mathbf{2},\mathbf 1 \oplus \mathbf 1}^{\mathrm{twi},(2)}(A)
    =
    \log\frac{8}{3}\, ,
\end{align}
for the generalized cluster state, whereas both quantities vanish for the product state \eqref{eq:trivial}. The twisted entropic order parameter thus distinguishes the $S_3\times\operatorname{Rep}(S_3)$ generalized cluster state from the trivial product state, and the reference representations for which it vanishes identify the $S_3$ representations carried by the defect endpoints.

It would be interesting to test these predictions numerically away from the fixed points, for instance in the interpolating model $H(\lambda)=(1-\lambda)H_{\rm tri}+\lambda H_{\rm cluster}$, which is symmetric under $G\times\operatorname{Rep}(G)$ for all $\lambda$ and reduces to the $N=2$ model of section~\ref{sec:example} for $G=\mathbb Z_2$. Since the two fixed-point Hamiltonians do not commute, this model is genuinely interacting for $0<\lambda<1$, and the twisted entropic order parameter is expected to interpolate between the two fixed-point values across the transition, in the same way as in figure~\ref{fig:Z2_n1_plot}. We leave this to future work.


\section{Conclusion and future directions}\label{sec:conclusion}
In this work, we introduced the twisted entropic order parameter as a density-matrix probe of symmetry-defect endpoint charges and applied it to one-dimensional bosonic SPT phases.

In section~\ref{sec:tea}, we formulated the general construction for a symmetry-broken phase $G\to H$. The essential step is not merely to introduce an ancilla qubit that coherently records the untwisted and twisted branches, but also to endow the ancilla with a symmetry action twisted by the scanning character $\lambda$ whose charge is being tested. This character-dependent coupling allows the corresponding twirling channel to project the transition operator onto a definite endpoint-charge sector. We showed that scanning over $\lambda$ identifies the physical endpoint character determined by the projective edge representation. We also introduced the R\'enyi generalization and derived a simple expression for the second R\'enyi twisted entropic order parameter in terms of the symmetry-resolved coherence between the two branches. We also studied the clock-broken $\mathbb Z_N$ cluster ladder model. As a demonstration, we analytically identified the endpoint charges carried by the partial symmetry strings in the different SPT phases and showed that the twisted entropic order parameter distinguishes the corresponding $\mathbb Z_N$-valued SPT indices. We also numerically investigated its behavior as the coupling constants were varied. In particular, deep inside each SPT phase, the character scan correctly selects the localized endpoint charge expected from the corresponding SPT index, while its evolution away from the fixed-point limit 
provides evidence that the diagnostic remains effective away from the exactly solvable point within this decoupled model.

In section~\ref{sec:repG}, we considered $\operatorname{Rep}(G)$ symmetry as a representative example of a non-invertible symmetry. For the generalized cluster state protected by $G\times\operatorname{Rep}(G)$, we formulated an analogous twisted entropic construction and analyzed the $G$-charge carried by the endpoint of an open $\operatorname{Rep}(G)$ symmetry operator. We found that the resulting twisted entropic order parameter distinguishes the generalized cluster state from the corresponding product-state phase. This example suggests that the defect-endpoint approach underlying our construction can remain useful in systems with non-invertible symmetries.

Several interesting directions remain for future investigation.

\paragraph{Detecting SPT phases of mixed states.}
An important direction of future work is to generalize the twisted entropic order parameter to SPT phases of mixed states. Mixed-state SPT phases extend the notion of pure-state SPT phases to density matrices. Several related formulations have been proposed, including phase equivalence under symmetry-preserving local quantum channels \cite{Ma:2022pvq, Coser2019classificationof, Sang:2023rsp} and characterization based on mapping the density matrix to a pure state in a doubled Hilbert space \cite{Ma:2024kma, Xue:2024bkt}. Moreover, mixed states admit distinct notions of strong and weak symmetry \cite{deGroot:2021vdi,Ma:2022pvq}. It would therefore be interesting to investigate whether the twisted entropic order parameter can provide a unified diagnostic of mixed-state SPT phases.

\paragraph{Detecting symmetry fractionalization classes of SET phases.} 
An interesting future direction is to extend our framework to symmetry-enriched topological (SET) phases. Topologically ordered states can be viewed as phases with spontaneously broken one-form symmetry, and they can therefore be diagnosed by entanglement asymmetry associated with the one-form symmetry~\cite{Lamas:2025eay, Benini:2025hbj}. SET phases are further characterized by symmetry fractionalization classes $H^2_{\rho}(G,\mathcal{A})$, where $\mathcal{A}$ denotes the group of Abelian anyons and $\rho$ specifies the action of the zero-form global symmetry $G$ on the anyons~\cite{Barkeshli:2014cna}. These classes describe how the global symmetry $G$ is fractionalized on anyonic excitations. It would therefore be interesting to investigate whether these symmetry fractionalization classes can also be detected by the entropy-like order parameter.

\paragraph{Generalization to higher-dimensional SPT phases.}
It would be interesting to extend the present construction to SPT phases in higher spatial dimensions; see e.g.~\cite{Chen:2011pg, Levin:2012yb,Vishwanath:2012tq}. In this case, the boundary of a partially inserted symmetry defect is no longer pointlike, but forms an extended object along the entangling surface. A natural question is whether an appropriately generalized twisted entropic order parameter can extract the lower-dimensional symmetry response carried by this defect boundary.

\paragraph{Fermionic SPT phases.}
Another important direction is the extension to fermionic SPT phases; see e.g.~\cite{Kitaev:2000nmw,Fidkowski_2010,Fidkowski_2011,Gu:2012ib, Kitaev2009, Schnyder:2008tya, Kapustin:2014dxa, Gaiotto:2015zta, Ryu:2010zza}. Such a generalization must incorporate fermion-parity superselection and the graded structure of fermionic Hilbert spaces into both the ancilla construction and the symmetry-twirling operation. It would be particularly interesting to determine whether the resulting quantity can detect fermionic edge data, such as Majorana boundary zero-modes, directly from reduced density matrices.

\paragraph{Anomaly-free fusion-category SPT phases.}
Fusion-category symmetries and the gapped phases protected by them provide a natural generalization of ordinary group-symmetric phases; see, e.g.~\cite{Thorngren:2019iar, Inamura:2021szw, Inamura:2021wuo, Inamura:2024jke, Seifnashri:2024dsd, Li:2024fhy, Maeda:2025rxc, Lu:2025yru, Cao:2025qhg, Lu:2025rwd, Aksoy:2025rmg, Pace:2024acq, Warman:2024lir}. In section~\ref{sec:repG}, we formulated the twisted entropic order parameter for $\operatorname{Rep}(G)$ symmetry. More generally, an anomaly-free unitary fusion-category symmetry is specified by a fusion category $\mathcal C$ together with a fiber functor $f$. By Tannaka reconstruction, such data $(\mathcal{C} , f)$ can be described by the representation category of a finite-dimensional Hopf $C^\ast$-algebra equipped with its forgetful functor, and admit an onsite MPO realization on a
tensor-product Hilbert space~\cite{Meng:2024nxx}. It would therefore be
interesting to generalize the construction of section~\ref{sec:repG} from
$\operatorname{Rep}(G)$ to this general setting. In particular, the virtual spaces of the Hopf-algebra MPO could be retained as auxiliary defect registers, allowing the twisted entropic order parameter to probe the charge carried by the endpoint of an open symmetry operator. Such a construction could provide a density-matrix diagnostic for a broad class of anomaly-free non-invertible SPT phases.

 \acknowledgments
 We thank Hiromi~Ebisu and Yuya~Tanizaki for helpful discussions and insightful comments on the draft.
 KF is supported by Grant-in-Aid for JSPS Fellows No.~26KJ1554.
 TO is supported by JST SPRING No.~JPMJSP2110.
 SS is supported by Grant-in-Aid for JSPS Fellows No.~26KJ0182. 

\appendix

\section{Upper bound of twisted entropic order parameter}
\label{sec:upperbound}

In this appendix, we derive the upper bound of twisted entropic order parameter~\eqref{eq:upperbound}.  We then
determine when this bound is saturated.  The origin of the bound is the two-branch structure of the ancilla construction and is independent of the order of the symmetry group. For notational simplification, let us introduce 
\begin{align}
    R_{CA}
    &\equiv
    \rho_{CA}^{(g)}
    =
    \frac{1}{2}
    \begin{pmatrix}
        \rho_A & X_g \\
        X_g^\dagger & \rho_A
    \end{pmatrix}_C\, ,
    \label{eq:app_R_CA}
    \\
    \sigma_{CA}^{(\lambda)}
    &\equiv
    \mathcal G_{g,\lambda}^{CA}
    \left(
        \rho_{CA}^{(g)}
    \right)
    =
    \frac{1}{2}
    \begin{pmatrix}
        \rho_A & X_{g,\lambda} \\
        X_{g,\lambda}^\dagger & \rho_A
    \end{pmatrix}_C\, .
    \label{eq:app_sigma_CA}
\end{align}
We introduce complete dephasing of the ancilla in the basis $\{\ket{0}_C,\ket{1}_C\}$,
\begin{align}\label{eq:app_ancilla_dephasing}
    \mathcal D_C(\mathcal O)
    \equiv
    \sum_{c=0,1}
    \Cketbra{c}\,
    \mathcal O\,
    \Cketbra{c}\, .
\end{align}
Since the diagonal blocks of both density matrices in
\eqref{eq:app_R_CA} and \eqref{eq:app_sigma_CA} are equal, their
dephased states coincide:
\begin{align}\label{eq:tau}
    \begin{aligned}
    \tau_{CA}
    &\equiv
    \mathcal D_C(R_{CA})
    =
    \mathcal D_C
    \left(
        \sigma_{CA}^{(\lambda)}
    \right) \\
    &=
    \frac{\mathbf 1_C}{2}
    \otimes
    \rho_A\, .
    \end{aligned}
\end{align}
The dephasing channel can be written as a binary random-unitary channel,
\begin{align}\label{eq:app_dephasing_random_unitary}
    \mathcal D_C(\mathcal O)
    =
    \frac{1}{2}
    \left(
    \mathcal{O}+Z_{C} \mathcal{O} Z_{C} 
    \right)\, , \quad
    Z_{C}
    \equiv
    \Cketbra{0} - \Cketbra{1}\, . 
\end{align}
First, define
\begin{align}
    R_{0}
    \equiv
    R_{CA}\, , \quad
    R_{1}
    \equiv
    Z_C R_{CA} Z_C\, , 
\end{align}
then $\tau_{CA}$ defined in \eqref{eq:tau} can be rewritten as
\begin{align}\label{eq:tau_identity}
    \tau_{CA}
    =
    \frac{R_0 + R_1 }{2}\, . 
\end{align}

In terms of these new variables, the twisted entropic order parameter admits the following decomposition,
\begin{align}
    \Delta S_{g,\lambda}^{\mathrm{twi}}(A)
    =
    S_{CA}(\tau_{CA})
    -
    S_{CA}(R_{CA})
    -
    D\left(\sigma_{CA}^{(\lambda)}\Vert \tau_{CA} \right)\, . 
\end{align}
Therefore, by using the non-negativity of relative entropy, one can obtain the following inequality, 
\begin{align}\label{eq:tau_upper}
    \Delta S_{g,\lambda}^{\mathrm{twi}}(A)
    \leq 
    S_{CA}(\tau_{CA})
    -
    S_{CA}(R_{CA})\, ,
\end{align}
with equality precisely when $\sigma_{CA}^{(\lambda)} = \tau_{CA}$.

Expanding the relative entropies gives the exact identity
\begin{align}\label{eq:app_binary_mixing_identity}
    S_{CA}(\tau_{CA})
    -
    \frac{1}{2}
    \sum_{j=0}^{1}S_{CA}(R_j)
    &=
    \frac{1}{2}
    \sum_{j=0}^{1}
    D(R_j\Vert\tau_{CA})\, .
\end{align}
Indeed, thanks to \eqref{eq:tau_identity}, the terms containing $\log\tau_{CA}$ combine according to
\begin{align}
    \frac{1}{2}
    \sum_{j=0}^{1}
    \Tr_{CA}
    \left(
        R_j\log\tau_{CA}
    \right)
    =
    \Tr_{CA}
    \left(
        \tau_{CA}\log\tau_{CA}
    \right)\, .
\end{align}
On the other hand, the following relation also holds,
\begin{align}
    \tau_{CA}
    -
    \frac{R_j}{2}
    =
    \frac{R_{1-j}}{2} \, , 
\end{align}
which implies that the operator in the left-hand side is positive semidefinite. Namely, 
\begin{align}
    \mathcal{M}
    \succeq 0\, , \quad 
    \mathcal{M}_{\phi \varphi}
    \equiv
    \bra{\phi}
    \left(
    \tau_{CA}
    -
    \frac{R_j}{2}
    \right)
    \ket{\varphi}\, .
\end{align}
By the operator monotonicity of logarithm, this gives
\begin{align}
    \log \tau_{CA}
    \succeq 
    \log \frac{R_{j}}{2}
    =
    \log R_j 
    -
    (\log 2)\, \mathbf{1}\, . 
\end{align}
Substitution into \eqref{eq:app_binary_mixing_identity} yields
\begin{align}\label{eq:app_binary_mixing_bound}
    S_{CA}(\tau_{CA})
    &\leq
    \log 2
    +
    \frac{1}{2}S_{CA}(R_0)
    +
    \frac{1}{2}S_{CA}(R_1)
    \nonumber\\
    &=
    S_{CA}(R_{CA})
    +
    \log 2\, ,
\end{align}
To verify the last equality explicitly, recall that $Z_C^\dagger=Z_C$ and $Z_C^2=\mathbf 1_C$. Hence
$R_1=Z_C R_0 Z_C^\dagger$ is unitarily equivalent to $R_0$, and
\begin{align}
    \log R_1
    =
    Z_C(\log R_0)Z_C^\dagger\, .
\end{align}
Using cyclicity of the trace, we then find
\begin{align}\label{eq:app_unitary_entropy_invariance}
    S_{CA}(R_1)
    &=
    -\Tr_{CA}
    \left(
        R_1\log R_1
    \right)
    \nonumber\\
    &=
    -\Tr_{CA}
    \left[
        Z_C
        \left(
            R_0\log R_0
        \right)
        Z_C^\dagger
    \right]
    =
    S_{CA}(R_0)
    =
    S_{CA}(R_{CA})\, .
\end{align}
Combining \eqref{eq:tau_upper} with \eqref{eq:app_binary_mixing_bound}, we obtain
\begin{align}
    \Delta S_{g,\lambda}^{\mathrm{twi}}(A)
    &\leq
    S_{CA}(\tau_{CA})
    -
    S_{CA}(R_{CA})
    \leq
    \log 2\, .
    \label{eq:app_TEOP_upper_bound_proof}
\end{align}
We finally show that the upper bound is saturated at an SPT fixed point.  We restrict ourselves to a mismatched scanning character, $\lambda\neq\epsilon_g^\omega$, and suppress the fixed-point superscript on $\rho_A$ for notational simplicity.  All operator identities below are understood on the Schmidt support of $\rho_A$.
It follows that, for every real $n>0$,
\begin{align}
    \mathrm{Tr}_{CA}\left[\left(\mathcal G_{g,\lambda}^{CA}\left(\rho_{CA}^{(g)}\right)\right)^n\right]=\mathrm{Tr}_{CA}
    \frac{1}{2^n}
    \begin{pmatrix}
        \rho_A^n & 0
        \\
        0 & \rho_A^n
    \end{pmatrix}_C
    =\frac{1}{2^{n-1}}\mathrm{Tr}_A\rho_A^n\, . 
\end{align}
On the other hand, we obtain
\begin{align}
    \mathrm{Tr}_{CA}\left[\left(\rho_{CA}^{(g)}\right)^n\right]
    =\mathrm{Tr}_A\rho_A^n\, , 
\end{align}
where we have used the fixed point identities
\begin{align}
    \rho_AV_g=V_g\rho_A \, , \quad V_gV_g^{\dagger}=1\, \quad \text{at fixed points}\, . 
\end{align}
From the definition of the R\'enyi twisted entropic order parameter~\eqref{eq:Renyi_twisted_entropic_order_parameter_ratio}, we finally obtain 
\begin{align}
    \Delta S_{g,\lambda}^{\mathrm{twi},(n)}(A)=\frac{1}{1-n}\log\frac{\frac{1}{2^{n-1}}\mathrm{Tr}_A\rho_A^n}{\mathrm{Tr}_A\rho_A^n}=\log 2\, . 
\end{align}
Since it is independent of R\'enyi index $n$, we can safely take the $n\rightarrow1$ limit and obtain
\begin{align}
    \Delta S_{g,\lambda}^{\mathrm{twi}}(A)=\log 2\, ,
\end{align}
which is precisely when the inequality \eqref{eq:app_TEOP_upper_bound_proof} saturates.\footnote{Of course, one can arrive at the same conclusion without resorting to the R\'enyi analysis.}

\section{Derivation of \eqref{eq:RepG_matched_zero}}
\label{sec:derivation}

The aim of this appendix is to derive the formula \eqref{eq:RepG_matched_zero}. To this end, we first note that at the generalized-cluster fixed point,\footnote{Here, we omit the dependence on the reference vector $v$ for simplicity.}
\begin{align}
    \rho_{A,\Gamma;\alpha\beta}
    =
    V^\dagger_{\Gamma,\alpha}
    \rho_A
    V_{\Gamma,\beta} \, .
\end{align}
This identity can be shown as follows.
\begin{align}
    \begin{aligned}
    \rho_{A,\Gamma;\alpha\beta}
    &=
    \Tr_B
    \left[
        R_{\Gamma,\alpha}^{B}
        \ket{\rm cluster}\bra{\rm cluster}
        R_{\Gamma,\beta}^{B\dagger}
    \right] \\
    &=
    \Tr_B
    \left[
        V_{\Gamma,\alpha}^\dagger
        \ket{\rm cluster}\bra{\rm cluster}
        V_{\Gamma,\beta}
    \right] \\
    &=
    V_{\Gamma,\alpha}^\dagger
    \Tr_B
    \left[
        \ket{\rm cluster}\bra{\rm cluster}
    \right]
    V_{\Gamma,\beta}\\
    &=
    V_{\Gamma,\alpha}^\dagger
    \rho_A
    V_{\Gamma,\beta}\, ,
    \end{aligned}
\end{align}
where in the second line, we used the formula~\eqref{eq:action_R}. Also, in the third line, we used the fact that $V_{\Gamma, \alpha}$ acts only on $A$, so it can be taken outside the partial trace over $B$.
Moreover, by using the following properties,
\begin{align}
    U_g^A\rho_AU_g^{A\dagger}
    =
    \rho_A\, , \quad
    U_g^A
    V_{\Gamma,\alpha}
    U_g^{A\dagger}
    =
    \sum_{\mu=1}^{d_\Gamma}
    [\overline\Gamma(g)]_{\mu\alpha}
    V_{\Gamma,\mu}\, , 
\end{align}
we obtain
\begin{align}\label{eq:RepG_diagonal_covariance}
    U_g^A
    \rho_{A,\Gamma;\alpha\beta}
    U_g^{A\dagger}
    =
    \sum_{\mu,\nu=1}^{d_\Gamma}
        [\Gamma(g)]_{\mu\alpha}
    [\overline\Gamma(g)]_{\nu\beta}
    \rho_{A,\Gamma;\mu\nu}\, . 
\end{align}
When the representation is chosen as $\Lambda_{I_\star}=\Gamma_X$, the upper-right block of~\eqref{eq:RepG_twirled_density_matrix} transforms as
\begin{align}
    \begin{aligned}
    \sum_{\alpha=1}^{d_\Gamma}
        [\Gamma(g)]_{\beta\alpha}
    U_g^A
    X_{\Gamma,\alpha}
    U_g^{A\dagger}
    =
    \sum_{\alpha,\mu=1}^{d_\Gamma}
        [\Gamma(g)]_{\beta\alpha}
    [\overline\Gamma(g)]_{\mu\alpha}
    X_{\Gamma,\mu}
    =
    X_{\Gamma,\beta}\, . 
    \end{aligned}
\end{align}
where the unitarity of the matrix $\Gamma(g)$ was used. We note that $\Gamma_X=\overline\Gamma$ at the generalized-cluster fixed point as shown in \eqref{eq:RepG_fixed_point_endpoint_representations}. Similarly, the lower-right block of~\eqref{eq:RepG_twirled_density_matrix} reduces as 
\begin{align}
    \begin{aligned}
    &\sum_{\alpha,\beta=1}^{d_\Gamma}
    [\overline\Gamma(g)]_{\mu\alpha}
        [\Gamma(g)]_{\nu\beta}
    U_g^A
    \rho_{A,\Gamma;\alpha\beta}
    U_g^{A\dagger} \\
    &=
    \sum_{\alpha,\beta, \gamma, \varepsilon=1}^{d_\Gamma}
    [\overline\Gamma(g)]_{\mu\alpha}
        [\Gamma(g)]_{\nu\beta}[\Gamma(g)]_{\gamma\alpha}
    [\overline\Gamma(g)]_{\varepsilon \beta}
    \rho_{A,\Gamma;\gamma \varepsilon}
    \\
    &=
    \rho_{A,\Gamma;\mu\nu}\, . 
    \end{aligned}
\end{align}
Thus every block of the enlarged density matrix is invariant:
\begin{align}
    \widetilde U_g^{(\Gamma_X,I_\star)}
    \rho_{CA}^{(\Gamma)}
    \widetilde U_g^{(\Gamma_X,I_\star)\dagger}
    =
    \rho_{CA}^{(\Gamma)}
    \qquad
    \forall g\in G.
\end{align}
Averaging over $G$ therefore gives
\begin{align}
    \mathcal G_{\Gamma,\Gamma_X;I_\star}^{CA}
    \left(
        \rho_{CA}^{(\Gamma)}
    \right)
    =
    \rho_{CA}^{(\Gamma)}\, . 
\end{align}
By the faithfulness of the quantum relative entropy, it follows that
\begin{align}
    \Delta S_{\Gamma,\Gamma_X}^{\mathrm{twi}}(A;I_\star)
    =
    0\, . 
\end{align}

\section{Path integral interpretation of endpoint charge} \label{sec:field_theory}
In the main text, we have discussed the endpoint charge that appears at the endpoint of an open symmetry operator in an SPT phase. In this appendix, we briefly review a Euclidean path integral interpretation of this endpoint charge. In particular, we show that it is encoded by the slant product of the $2$-cocycle defining the SPT response, or equivalently by the symmetry-twisted torus partition function.

Let us consider a $G$-SPT phase on a torus characterized by a cohomology class $\omega\in H^2(G,\mathrm{U}(1))$. Its partition function in the presence of a background $G$ gauge field $A$ is given by
\begin{align}
Z_{\mathrm{SPT}}[A]
=
\exp\left(
2\pi \i \int_{T^2} A^{*}\omega
\right),
\end{align}
where $A^{*}\omega$ denotes the pullback of $\omega$ by $A$.
In the Euclidean spacetime picture, an open $g$-symmetry operator with an endpoint traces out a $g$-symmetry defect line extending along the time direction. We therefore fix $g\in G$ and consider the theory in the presence of a time-like $g$-symmetry defect. Equivalently, quantizing along the time direction gives the $g$-twisted Hilbert space $\mathcal H_g$.

To determine the symmetry charge carried by the $g$-defect, we further act with an element $h\in C_G(g)$, where $C_G(g)$ is the centralizer of $g$. In the Euclidean path-integral description, this amounts to inserting an $h$-symmetry defect along the spatial direction. If $\ket{\Psi_g}\in\mathcal H_g$ denotes the SPT state associated with the $g$-defect, the action of $h$ is characterized by
\begin{align}
U_h \ket{\Psi_g}
=
i_g\omega(h)\ket{\Psi_g}, \quad i_g\omega(h)
=
\frac{\omega(h,g)}{\omega(g,h)}\, ,
\end{align}
where $i_g\omega$ is the slant product of $\omega$ by $g$. This shows that inserting a $g$-defect along the time direction induces the dimensional reduction of SPT cocycle $\omega$ to the slant product~\cite{Tantivasadakarn:2017xbg}. From this relation, the torus partition function in the presence of the time-like $g$-symmetry defect and the spatial $h$-symmetry defect is given by\footnote{
More precisely, the partition function is defined as
\begin{align}
Z_{\mathrm{SPT}}[g,h]
=
\Tr_{\mathcal H_g}
\!\left(
U_h e^{-\beta H}
\right)\, .  
\end{align}
In the low-energy limit $\beta\to\infty$, assuming a unique ground state
$\ket{\Psi_g}$ and setting its ground-state energy to zero, the partition function reduces to
\begin{align}
Z_{\mathrm{SPT}}[g,h]
=
\bra{\Psi_g}U_h\ket{\Psi_g}.
\end{align}
}
\begin{align}
Z_{\mathrm{SPT}}[g,h]
= \bra{\Psi_g}U_h\ket{\Psi_g}
=i_g\omega(h)\, .
\end{align}
Thus, dimensional reduction of the SPT response in the presence of a $g$-symmetry defect produces the character $i_g\omega$. In particular, $i_g\omega(h)$ gives the $h$-charge carried by the time-like $g$-symmetry defect. Since this defect can be understood as the endpoint of an open $g$-symmetry operator, this charge is precisely the endpoint charge discussed in the main text and appearing in~\eqref{eq:Xg_endpoint_character_general}.

\bibliographystyle{JHEP}
\bibliography{refs}

\providecommand{\href}[2]{#2}\begingroup\raggedright\begin{thebibliography}{100}

\bibitem{Landau:1937obd}
L.~D. Landau, \emph{{On the theory of phase transitions}}, \href{https://doi.org/10.1016/B978-0-08-010586-4.50034-1}{\emph{Zh. Eksp. Teor. Fiz.} {\bfseries 7} (1937) 19}.

\bibitem{Ares:2022koq}
F.~Ares, S.~Murciano and P.~Calabrese, \emph{{Entanglement asymmetry as a probe of symmetry breaking}}, \href{https://doi.org/10.1038/s41467-023-37747-8}{\emph{Nature Commun.} {\bfseries 14} (2023) 2036} [\href{https://arxiv.org/abs/2207.14693}{{\ttfamily 2207.14693}}].

\bibitem{Capizzi:2023xaf}
L.~Capizzi and V.~Vitale, \emph{{A universal formula for the entanglement asymmetry of matrix product states}}, \href{https://doi.org/10.1088/1751-8121/ad8796}{\emph{J. Phys. A} {\bfseries 57} (2024) 45LT01} [\href{https://arxiv.org/abs/2310.01962}{{\ttfamily 2310.01962}}].

\bibitem{Ares:2023kcz}
F.~Ares, S.~Murciano, E.~Vernier and P.~Calabrese, \emph{{Lack of symmetry restoration after a quantum quench: An entanglement asymmetry study}}, \href{https://doi.org/10.21468/SciPostPhys.15.3.089}{\emph{SciPost Phys.} {\bfseries 15} (2023) 089} [\href{https://arxiv.org/abs/2302.03330}{{\ttfamily 2302.03330}}].

\bibitem{Rylands:2023yzx}
C.~Rylands, K.~Klobas, F.~Ares, P.~Calabrese, S.~Murciano and B.~Bertini, \emph{{Microscopic Origin of the Quantum Mpemba Effect in Integrable Systems}}, \href{https://doi.org/10.1103/PhysRevLett.133.010401}{\emph{Phys. Rev. Lett.} {\bfseries 133} (2024) 010401} [\href{https://arxiv.org/abs/2310.04419}{{\ttfamily 2310.04419}}].

\bibitem{Murciano:2023qrv}
S.~Murciano, F.~Ares, I.~Klich and P.~Calabrese, \emph{{Entanglement asymmetry and quantum Mpemba effect in the XY spin chain}}, \href{https://doi.org/10.1088/1742-5468/ad17b4}{\emph{J. Stat. Mech.} {\bfseries 2401} (2024) 013103} [\href{https://arxiv.org/abs/2310.07513}{{\ttfamily 2310.07513}}].

\bibitem{Yamashika:2024hpr}
S.~Yamashika, F.~Ares and P.~Calabrese, \emph{{Entanglement asymmetry and quantum Mpemba effect in two-dimensional free-fermion systems}}, \href{https://doi.org/10.1103/PhysRevB.110.085126}{\emph{Phys. Rev. B} {\bfseries 110} (2024) 085126} [\href{https://arxiv.org/abs/2403.04486}{{\ttfamily 2403.04486}}].

\bibitem{Chalas:2024wjz}
K.~Chalas, F.~Ares, C.~Rylands and P.~Calabrese, \emph{{Multiple crossings during dynamical symmetry restoration and implications for the quantum Mpemba effect}}, \href{https://doi.org/10.1088/1742-5468/ad769c}{\emph{J. Stat. Mech.} {\bfseries 2024} (2024) 103101} [\href{https://arxiv.org/abs/2405.04436}{{\ttfamily 2405.04436}}].

\bibitem{Benini:2024xjv}
F.~Benini, V.~Godet and A.~H. Singh, \emph{{Entanglement asymmetry in conformal field theory and holography}},  \href{https://arxiv.org/abs/2407.07969}{{\ttfamily 2407.07969}}.

\bibitem{Fujimura:2025rnm}
H.~Fujimura and S.~Shimamori, \emph{{Entanglement asymmetry and quantum Mpemba effect for non-Abelian global symmetry}}, \href{https://doi.org/10.1007/JHEP03(2026)244}{\emph{JHEP} {\bfseries 03} (2026) 244} [\href{https://arxiv.org/abs/2509.05597}{{\ttfamily 2509.05597}}].

\bibitem{Ares:2025onj}
F.~Ares, P.~Calabrese and S.~Murciano, \emph{{The quantum Mpemba effects}},  \href{https://arxiv.org/abs/2502.08087}{{\ttfamily 2502.08087}}.

\bibitem{Chen:2023gql}
M.~Chen and H.-H. Chen, \emph{{R{\'e}nyi entanglement asymmetry in (1+1)-dimensional conformal field theories}}, \href{https://doi.org/10.1103/PhysRevD.109.065009}{\emph{Phys. Rev. D} {\bfseries 109} (2024) 065009} [\href{https://arxiv.org/abs/2310.15480}{{\ttfamily 2310.15480}}].

\bibitem{Fossati:2024xtn}
M.~Fossati, F.~Ares, J.~Dubail and P.~Calabrese, \emph{{Entanglement asymmetry in CFT and its relation to non-topological defects}}, \href{https://doi.org/10.1007/JHEP05(2024)059}{\emph{JHEP} {\bfseries 05} (2024) 059} [\href{https://arxiv.org/abs/2402.03446}{{\ttfamily 2402.03446}}].

\bibitem{Kusuki:2024gss}
Y.~Kusuki, S.~Murciano, H.~Ooguri and S.~Pal, \emph{{Entanglement asymmetry and symmetry defects in boundary conformal field theory}}, \href{https://doi.org/10.1007/JHEP01(2025)057}{\emph{JHEP} {\bfseries 01} (2025) 057} [\href{https://arxiv.org/abs/2411.09792}{{\ttfamily 2411.09792}}].

\bibitem{Fossati:2024ekt}
M.~Fossati, C.~Rylands and P.~Calabrese, \emph{{Entanglement asymmetry in CFT with boundary symmetry breaking}}, \href{https://doi.org/10.1007/JHEP06(2025)089}{\emph{JHEP} {\bfseries 06} (2025) 089} [\href{https://arxiv.org/abs/2411.10244}{{\ttfamily 2411.10244}}].

\bibitem{Lastres:2024ohf}
M.~Lastres, S.~Murciano, F.~Ares and P.~Calabrese, \emph{{Entanglement asymmetry in the critical XXZ spin chain}}, \href{https://doi.org/10.1088/1742-5468/ada497}{\emph{J. Stat. Mech.} {\bfseries 2025} (2025) 013107} [\href{https://arxiv.org/abs/2407.06427}{{\ttfamily 2407.06427}}].

\bibitem{Fossati:2026jww}
M.~Fossati, C.~Rylands, E.~Grosfeld, E.~Sela and P.~Calabrese, \emph{{Boundary quenches in (1+1)-dimensional conformal field theory}},  \href{https://arxiv.org/abs/2607.19166}{{\ttfamily 2607.19166}}.

\bibitem{Chen:2024lxe}
H.-H. Chen and Z.-J. Tang, \emph{{Entanglement asymmetry in the Hayden-Preskill protocol}}, \href{https://doi.org/10.1103/PhysRevD.111.066003}{\emph{Phys. Rev. D} {\bfseries 111} (2025) 066003} [\href{https://arxiv.org/abs/2411.17695}{{\ttfamily 2411.17695}}].

\bibitem{Caceffo:2024jbc}
F.~Caceffo, S.~Murciano and V.~Alba, \emph{{Entangled multiplets, asymmetry, and quantum Mpemba effect in dissipative systems}}, \href{https://doi.org/10.1088/1742-5468/ad4537}{\emph{J. Stat. Mech.} {\bfseries 2024} (2024) 063103} [\href{https://arxiv.org/abs/2402.02918}{{\ttfamily 2402.02918}}].

\bibitem{Benini:2025lav}
F.~Benini, P.~Calabrese, M.~Fossati, A.~H. Singh and M.~Venuti, \emph{{Entanglement asymmetry for higher and noninvertible symmetries}},  \href{https://arxiv.org/abs/2509.16311}{{\ttfamily 2509.16311}}.

\bibitem{AliAhmad:2025bnd}
S.~Ali~Ahmad, M.~S. Klinger and Y.~Wang, \emph{{The many faces of non-invertible symmetries}}, \href{https://doi.org/10.1007/JHEP05(2026)110}{\emph{JHEP} {\bfseries 05} (2026) 110} [\href{https://arxiv.org/abs/2509.18072}{{\ttfamily 2509.18072}}].

\bibitem{Benini:2025hbj}
F.~Benini, E.~Garc{\'\i}a-Valdecasas and S.~Vitouladitis, \emph{{Higher-form entanglement asymmetry. Part I. The limits of symmetry breaking}}, \href{https://doi.org/10.1007/JHEP05(2026)202}{\emph{JHEP} {\bfseries 05} (2026) 202} [\href{https://arxiv.org/abs/2512.15898}{{\ttfamily 2512.15898}}].

\bibitem{Lamas:2025eay}
A.~G. Lamas, J.~Gliozzi and T.~L. Hughes, \emph{{Higher-form entanglement asymmetry and topological order}},  \href{https://arxiv.org/abs/2510.03967}{{\ttfamily 2510.03967}}.

\bibitem{Vescovo:2026okg}
M.~Vescovo, P.~Calabrese and F.~Ares, \emph{{Entanglement asymmetry and quantum Mpemba effect for Kramers-Wannier duality}},  \href{https://arxiv.org/abs/2607.21226}{{\ttfamily 2607.21226}}.

\bibitem{Gotta:2026tum}
L.~Gotta, F.~Ares and S.~Murciano, \emph{{Enhancing entanglement asymmetry in fragmented quantum systems}},  \href{https://arxiv.org/abs/2603.02338}{{\ttfamily 2603.02338}}.

\bibitem{Zhang:2020bpf}
H.-C. Zhang, G.~Sierra and J.~Molina-Vilaplana, \emph{{Entropic order parameters and topological holography}}, \href{https://doi.org/10.1007/JHEP06(2026)083}{\emph{JHEP} {\bfseries 26} (2020) 083} [\href{https://arxiv.org/abs/2512.24225}{{\ttfamily 2512.24225}}].

\bibitem{Gu:2009dr}
Z.-C. Gu and X.-G. Wen, \emph{{Tensor-Entanglement-Filtering Renormalization Approach and Symmetry Protected Topological Order}}, \href{https://doi.org/10.1103/PhysRevB.80.155131}{\emph{Phys. Rev. B} {\bfseries 80} (2009) 155131} [\href{https://arxiv.org/abs/0903.1069}{{\ttfamily 0903.1069}}].

\bibitem{Pollmann:2009ryx}
F.~Pollmann, A.~M. Turner, E.~Berg and M.~Oshikawa, \emph{{Entanglement spectrum of a topological phase in one dimension}}, \href{https://doi.org/10.1103/PhysRevB.81.064439}{\emph{Phys. Rev. B} {\bfseries 81} (2010) 064439} [\href{https://arxiv.org/abs/0910.1811}{{\ttfamily 0910.1811}}].

\bibitem{Chen:2010zpc}
X.~Chen, Z.-C. Gu and X.-G. Wen, \emph{{Classification of gapped symmetric phases in one-dimensional spin systems}}, \href{https://doi.org/10.1103/PhysRevB.83.035107}{\emph{Phys. Rev. B} {\bfseries 83} (2011) 035107} [\href{https://arxiv.org/abs/1008.3745}{{\ttfamily 1008.3745}}].

\bibitem{Chen:2011pg}
X.~Chen, Z.-C. Gu, Z.-X. Liu and X.-G. Wen, \emph{{Symmetry protected topological orders and the group cohomology of their symmetry group}}, \href{https://doi.org/10.1103/PhysRevB.87.155114}{\emph{Phys. Rev. B} {\bfseries 87} (2013) 155114} [\href{https://arxiv.org/abs/1106.4772}{{\ttfamily 1106.4772}}].

\bibitem{Chen:2011hnt}
X.~Chen, Z.-C. Gu and X.-G. Wen, \emph{{Complete classification of one-dimensional gapped quantum phases in interacting spin systems}}, \href{https://doi.org/10.1103/PhysRevB.84.235128}{\emph{Phys. Rev. B} {\bfseries 84} (2011) 235128} [\href{https://arxiv.org/abs/1103.3323}{{\ttfamily 1103.3323}}].

\bibitem{Schuch:2011niz}
N.~Schuch, D.~P{\'e}rez-Garc{\'\i}a and I.~Cirac, \emph{{Classifying quantum phases using matrix product states and projected entangled pair states}}, \href{https://doi.org/10.1103/PhysRevB.84.165139}{\emph{Phys. Rev. B} {\bfseries 84} (2011) 165139} [\href{https://arxiv.org/abs/1010.3732}{{\ttfamily 1010.3732}}].

\bibitem{Levin:2012yb}
M.~Levin and Z.-C. Gu, \emph{{Braiding statistics approach to symmetry-protected topological phases}}, \href{https://doi.org/10.1103/PhysRevB.86.115109}{\emph{Phys. Rev. B} {\bfseries 86} (2012) 115109} [\href{https://arxiv.org/abs/1202.3120}{{\ttfamily 1202.3120}}].

\bibitem{Pollmann:2012}
F.~Pollmann, E.~Berg, A.~M. Turner and M.~Oshikawa, \emph{Symmetry protection of topological phases in one-dimensional quantum spin systems}, \href{https://doi.org/10.1103/PhysRevB.85.075125}{\emph{Phys. Rev. B} {\bfseries 85} (2012) 075125} [\href{https://arxiv.org/abs/0909.4059}{{\ttfamily 0909.4059}}].

\bibitem{Chen_2014}
X.~Chen, Y.-M. Lu and A.~Vishwanath, \emph{Symmetry-protected topological phases from decorated domain walls}, \href{https://doi.org/10.1038/ncomms4507}{\emph{Nature Communications} {\bfseries 5} (2014) }.

\bibitem{Pollmann2010EntanglementSpectrum}
F.~Pollmann, A.~M. Turner, E.~Berg and M.~Oshikawa, \emph{Entanglement spectrum of a topological phase in one dimension}, \href{https://doi.org/10.1103/PhysRevB.81.064439}{\emph{Phys. Rev. B} {\bfseries 81} (2010) 064439} [\href{https://arxiv.org/abs/0910.1811}{{\ttfamily 0910.1811}}].

\bibitem{Li2013IdentifyingSPT}
W.~Li, A.~Weichselbaum and J.~von Delft, \emph{Identifying symmetry-protected topological order by entanglement entropy}, \href{https://doi.org/10.1103/PhysRevB.88.245121}{\emph{Phys. Rev. B} {\bfseries 88} (2013) 245121} [\href{https://arxiv.org/abs/1306.5671}{{\ttfamily 1306.5671}}].

\bibitem{Marvian2017SPTEntanglement}
I.~Marvian, \emph{Symmetry-protected topological entanglement}, \href{https://doi.org/10.1103/PhysRevB.95.045111}{\emph{Phys. Rev. B} {\bfseries 95} (2017) 045111} [\href{https://arxiv.org/abs/1307.6617}{{\ttfamily 1307.6617}}].

\bibitem{Matsuura:2016qqu}
S.~Matsuura, X.~Wen, L.-Y. Hung and S.~Ryu, \emph{{Charged Topological Entanglement Entropy}}, \href{https://doi.org/10.1103/PhysRevB.93.195113}{\emph{Phys. Rev. B} {\bfseries 93} (2016) 195113} [\href{https://arxiv.org/abs/1601.03751}{{\ttfamily 1601.03751}}].

\bibitem{Azses2020SymmetryResolved}
D.~Azses and E.~Sela, \emph{Symmetry-resolved entanglement in symmetry-protected topological phases}, \href{https://doi.org/10.1103/PhysRevB.102.235157}{\emph{Phys. Rev. B} {\bfseries 102} (2020) 235157} [\href{https://arxiv.org/abs/2008.09332}{{\ttfamily 2008.09332}}].

\bibitem{Azses2023SymmetryResolved}
D.~Azses, D.~F. Mross and E.~Sela, \emph{Symmetry-resolved entanglement of two-dimensional symmetry-protected topological states}, \href{https://doi.org/10.1103/PhysRevB.107.115113}{\emph{Phys. Rev. B} {\bfseries 107} (2023) 115113} [\href{https://arxiv.org/abs/2210.12750}{{\ttfamily 2210.12750}}].

\bibitem{Shapourian2017ManyBody}
H.~Shapourian, K.~Shiozaki and S.~Ryu, \emph{Many-body topological invariants for fermionic symmetry-protected topological phases}, \href{https://doi.org/10.1103/PhysRevLett.118.216402}{\emph{Phys. Rev. Lett.} {\bfseries 118} (2017) 216402} [\href{https://arxiv.org/abs/1607.03896}{{\ttfamily 1607.03896}}].

\bibitem{Turzillo2025Detection}
A.~Turzillo, N.~Manjunath and J.~Garre-Rubio, \emph{Detection of two-dimensional symmetry protected topological order with partial symmetries}, \href{https://doi.org/10.1103/mfwg-svcp}{\emph{Phys. Rev. B} {\bfseries 112} (2025) 035119} [\href{https://arxiv.org/abs/2503.04510}{{\ttfamily 2503.04510}}].

\bibitem{Sala2026Entanglement}
P.~Sala, F.~Pollmann, M.~Oshikawa and Y.~You, \emph{Entanglement holography in quantum phases via twisted r{\'e}nyi-{N} correlators}, \href{https://doi.org/10.1088/2058-9565/ae4923}{\emph{Quantum Sci. Technol.} {\bfseries 11} (2026) 025017} [\href{https://arxiv.org/abs/2506.10076}{{\ttfamily 2506.10076}}].

\bibitem{Sohal:2026tpv}
R.~Sohal, M.~Levin and R.~Verresen, \emph{{Symmetry-Twisted Multi-Entropies: Order Parameters for 2D SPT Phases}},  \href{https://arxiv.org/abs/2607.12023}{{\ttfamily 2607.12023}}.

\bibitem{Pollmann_2012}
F.~Pollmann and A.~M. Turner, \emph{Detection of symmetry-protected topological phases in one dimension}, \href{https://doi.org/10.1103/physrevb.86.125441}{\emph{Physical Review B} {\bfseries 86} (2012) }.

\bibitem{Bogomolov1988}
F.~A. Bogomolov, \emph{The brauer group of quotient spaces by linear group actions}, \href{https://doi.org/10.1070/IM1988v030n03ABEH001024}{\emph{Mathematics of the USSR-Izvestiya} {\bfseries 30} (1988) 455}.

\bibitem{Moravec2012}
P.~Moravec, \emph{Unramified brauer groups of finite and infinite groups}, {\emph{American Journal of Mathematics} {\bfseries 134} (2012) 1679}.

\bibitem{Davydov:2013xov}
A.~Davydov, \emph{{Bogomolov multiplier, double class-preserving automorphisms, and modular invariants for orbifolds}}, \href{https://doi.org/10.1063/1.4895764}{\emph{J. Math. Phys.} {\bfseries 55} (2014) 092305} [\href{https://arxiv.org/abs/1312.7466}{{\ttfamily 1312.7466}}].

\bibitem{Kobayashi:2025pxs}
R.~Kobayashi and H.~Watanabe, \emph{{Projective Representations, Bogomolov Multiplier, and Their Applications in Physics}}, \href{https://doi.org/10.1007/s10773-025-06215-y}{\emph{Int. J. Theor. Phys.} {\bfseries 65} (2026) 65} [\href{https://arxiv.org/abs/2507.12515}{{\ttfamily 2507.12515}}].

\bibitem{Kobayashi:2025ykb}
R.~Kobayashi and M.~Barkeshli, \emph{{Soft symmetries of topological orders}}, \href{https://doi.org/10.1103/cwn4-jl57}{\emph{Phys. Rev. B} {\bfseries 113} (2026) 115150} [\href{https://arxiv.org/abs/2501.03314}{{\ttfamily 2501.03314}}].

\bibitem{Fechisin:2023odt}
C.~Fechisin, N.~Tantivasadakarn and V.~V. Albert, \emph{{Noninvertible Symmetry-Protected Topological Order in a Group-Based Cluster State}}, \href{https://doi.org/10.1103/PhysRevX.15.011058}{\emph{Phys. Rev. X} {\bfseries 15} (2025) 011058} [\href{https://arxiv.org/abs/2312.09272}{{\ttfamily 2312.09272}}].

\bibitem{Bhardwaj:2017xup}
L.~Bhardwaj and Y.~Tachikawa, \emph{{On finite symmetries and their gauging in two dimensions}}, \href{https://doi.org/10.1007/JHEP03(2018)189}{\emph{JHEP} {\bfseries 03} (2018) 189} [\href{https://arxiv.org/abs/1704.02330}{{\ttfamily 1704.02330}}].

\bibitem{Thorngren:2019iar}
R.~Thorngren and Y.~Wang, \emph{{Fusion category symmetry. Part I. Anomaly in-flow and gapped phases}}, \href{https://doi.org/10.1007/JHEP04(2024)132}{\emph{JHEP} {\bfseries 04} (2024) 132} [\href{https://arxiv.org/abs/1912.02817}{{\ttfamily 1912.02817}}].

\bibitem{Thorngren:2021yso}
R.~Thorngren and Y.~Wang, \emph{{Fusion category symmetry. Part II. Categoriosities at c = 1 and beyond}}, \href{https://doi.org/10.1007/JHEP07(2024)051}{\emph{JHEP} {\bfseries 07} (2024) 051} [\href{https://arxiv.org/abs/2106.12577}{{\ttfamily 2106.12577}}].

\bibitem{Zeng:2015pxf}
B.~Zeng, X.~Chen, D.-L. Zhou and X.-G. Wen, \emph{{Quantum Information Meets Quantum Matter: From Quantum Entanglement to Topological Phases of Many-Body Systems}}, Quantum Science and Technology. Springer, 2019, \href{https://doi.org/10.1007/978-1-4939-9084-9}{10.1007/978-1-4939-9084-9}, [\href{https://arxiv.org/abs/1508.02595}{{\ttfamily 1508.02595}}].

\bibitem{Tasaki:2020cpn}
H.~Tasaki, \emph{{Physics and Mathematics of Quantum Many-Body Systems}}. 2020, \href{https://doi.org/10.1007/978-3-030-41265-4}{10.1007/978-3-030-41265-4}.

\bibitem{Ogata2021Classification}
Y.~Ogata, \emph{A classification of pure states on quantum spin chains satisfying the split property with on-site finite group symmetries}, \href{https://doi.org/10.1090/btran/51}{\emph{Transactions of the American Mathematical Society, Series B} {\bfseries 8} (2021) 39}.

\bibitem{Else:2014vma}
D.~V. Else and C.~Nayak, \emph{{Classifying symmetry-protected topological phases through the anomalous action of the symmetry on the edge}}, \href{https://doi.org/10.1103/PhysRevB.90.235137}{\emph{Phys. Rev. B} {\bfseries 90} (2014) 235137} [\href{https://arxiv.org/abs/1409.5436}{{\ttfamily 1409.5436}}].

\bibitem{Gai:2026hjk}
Y.~Gai, S.~Schafer-Nameki and A.~Warman, \emph{{Twin Algebras: Condensable Algebras beyond Anyons}},  \href{https://arxiv.org/abs/2605.31602}{{\ttfamily 2605.31602}}.

\bibitem{Warman:2026gfz}
A.~Warman, Y.~Gai and S.~Schafer-Nameki, \emph{{Twin Phases: Intrinsic Deconfined Quantum Criticality}},  \href{https://arxiv.org/abs/2605.31601}{{\ttfamily 2605.31601}}.

\bibitem{Zhou:2003kna}
D.~L. Zhou, B.~Zeng, Z.~Xu and C.~P. Sun, \emph{{Quantum computation based on d-level cluster state}}, \href{https://doi.org/10.1103/PhysRevA.68.062303}{\emph{Phys. Rev. A} {\bfseries 68} (2003) 062303} [\href{https://arxiv.org/abs/quant-ph/0304054}{{\ttfamily quant-ph/0304054}}].

\bibitem{Brell:2015vqr}
C.~G. Brell, \emph{{Generalized Cluster States Based on Finite Groups}}, \href{https://doi.org/10.1088/1367-2630/17/2/023029}{\emph{New J. Phys.} {\bfseries 17} (2015) 023029} [\href{https://arxiv.org/abs/1408.6237}{{\ttfamily 1408.6237}}].

\bibitem{Pace:2024acq}
S.~D. Pace, H.~T. Lam and {\"O}.~M. Aksoy, \emph{{(SPT-)LSM theorems from projective non-invertible symmetries}}, \href{https://doi.org/10.21468/SciPostPhys.18.1.028}{\emph{SciPost Phys.} {\bfseries 18} (2025) 028} [\href{https://arxiv.org/abs/2409.18113}{{\ttfamily 2409.18113}}].

\bibitem{Albert:2021vts}
V.~V. Albert, D.~Aasen, W.~Xu, W.~Ji, J.~Alicea and J.~Preskill, \emph{{Spin chains, defects, and quantum wires for the quantum-double edge}},  \href{https://arxiv.org/abs/2111.12096}{{\ttfamily 2111.12096}}.

\bibitem{Ohyama:2026oay}
S.~Ohyama and K.~Inamura, \emph{{Parameterized families of 2+1d $G$-cluster states}},  \href{https://arxiv.org/abs/2601.08616}{{\ttfamily 2601.08616}}.

\bibitem{Inamura:2026hjl}
K.~Inamura and S.~Ohyama, \emph{{Generalized cluster states in 2+1d: non-invertible symmetries, interfaces, and parameterized families}},  \href{https://arxiv.org/abs/2601.08615}{{\ttfamily 2601.08615}}.

\bibitem{Jia:2024bng}
Z.~Jia, \emph{{Generalized cluster states from Hopf algebras: non-invertible symmetry and Hopf tensor network representation}}, \href{https://doi.org/10.1007/JHEP09(2024)147}{\emph{JHEP} {\bfseries 09} (2024) 147} [\href{https://arxiv.org/abs/2405.09277}{{\ttfamily 2405.09277}}].

\bibitem{wang2026classificationintrinsicallymixed11d}
Y.~Wang, \emph{Classification of intrinsically mixed $1+1$d non-invertible rep$(g) \times g$ spt phases},  \href{https://arxiv.org/abs/2603.24289}{{\ttfamily 2603.24289}}.

\bibitem{Seifnashri:2024dsd}
S.~Seifnashri and S.-H. Shao, \emph{{Cluster State as a Noninvertible Symmetry-Protected Topological Phase}}, \href{https://doi.org/10.1103/PhysRevLett.133.116601}{\emph{Phys. Rev. Lett.} {\bfseries 133} (2024) 116601} [\href{https://arxiv.org/abs/2404.01369}{{\ttfamily 2404.01369}}].

\bibitem{You:2025uxo}
M.~You, \emph{{Symmetric entanglers for non-invertible SPT phases}},  \href{https://arxiv.org/abs/2509.04581}{{\ttfamily 2509.04581}}.

\bibitem{Ma:2022pvq}
R.~Ma and C.~Wang, \emph{{Average Symmetry-Protected Topological Phases}}, \href{https://doi.org/10.1103/PhysRevX.13.031016}{\emph{Phys. Rev. X} {\bfseries 13} (2023) 031016} [\href{https://arxiv.org/abs/2209.02723}{{\ttfamily 2209.02723}}].

\bibitem{Coser2019classificationof}
A.~Coser and D.~P{\'{e}}rez-Garc{\'{i}}a, \emph{Classification of phases for mixed states via fast dissipative evolution}, \href{https://doi.org/10.22331/q-2019-08-12-174}{\emph{{Quantum}} {\bfseries 3} (2019) 174}.

\bibitem{Sang:2023rsp}
S.~Sang, Y.~Zou and T.~H. Hsieh, \emph{{Mixed-State Quantum Phases: Renormalization and Quantum Error Correction}}, \href{https://doi.org/10.1103/PhysRevX.14.031044}{\emph{Phys. Rev. X} {\bfseries 14} (2024) 031044} [\href{https://arxiv.org/abs/2310.08639}{{\ttfamily 2310.08639}}].

\bibitem{Ma:2024kma}
R.~Ma and A.~Turzillo, \emph{{Symmetry-Protected Topological Phases of Mixed States in the Doubled Space}}, \href{https://doi.org/10.1103/PRXQuantum.6.010348}{\emph{PRX Quantum} {\bfseries 6} (2025) 010348} [\href{https://arxiv.org/abs/2403.13280}{{\ttfamily 2403.13280}}].

\bibitem{Xue:2024bkt}
H.~Xue, J.~Y. Lee and Y.~Bao, \emph{{Tensor network formulation of symmetry protected topological phases in mixed states}},  \href{https://arxiv.org/abs/2403.17069}{{\ttfamily 2403.17069}}.

\bibitem{deGroot:2021vdi}
C.~de~Groot, A.~Turzillo and N.~Schuch, \emph{{Symmetry Protected Topological Order in Open Quantum Systems}}, \href{https://doi.org/10.22331/q-2022-11-10-856}{\emph{Quantum} {\bfseries 6} (2022) 856} [\href{https://arxiv.org/abs/2112.04483}{{\ttfamily 2112.04483}}].

\bibitem{Barkeshli:2014cna}
M.~Barkeshli, P.~Bonderson, M.~Cheng and Z.~Wang, \emph{{Symmetry Fractionalization, Defects, and Gauging of Topological Phases}}, \href{https://doi.org/10.1103/PhysRevB.100.115147}{\emph{Phys. Rev. B} {\bfseries 100} (2019) 115147} [\href{https://arxiv.org/abs/1410.4540}{{\ttfamily 1410.4540}}].

\bibitem{Vishwanath:2012tq}
A.~Vishwanath and T.~Senthil, \emph{{Physics of three dimensional bosonic topological insulators: Surface Deconfined Criticality and Quantized Magnetoelectric Effect}}, \href{https://doi.org/10.1103/PhysRevX.3.011016}{\emph{Phys. Rev. X} {\bfseries 3} (2013) 011016} [\href{https://arxiv.org/abs/1209.3058}{{\ttfamily 1209.3058}}].

\bibitem{Kitaev:2000nmw}
A.~Kitaev, \emph{{Unpaired Majorana fermions in quantum wires}}, \href{https://doi.org/10.1070/1063-7869/44/10S/S29}{\emph{Phys. Usp.} {\bfseries 44} (2001) 131} [\href{https://arxiv.org/abs/cond-mat/0010440}{{\ttfamily cond-mat/0010440}}].

\bibitem{Fidkowski_2010}
L.~Fidkowski and A.~Kitaev, \emph{Effects of interactions on the topological classification of free fermion systems}, \href{https://doi.org/10.1103/physrevb.81.134509}{\emph{Physical Review B} {\bfseries 81} (2010) }.

\bibitem{Fidkowski_2011}
L.~Fidkowski and A.~Kitaev, \emph{Topological phases of fermions in one dimension}, \href{https://doi.org/10.1103/physrevb.83.075103}{\emph{Physical Review B} {\bfseries 83} (2011) }.

\bibitem{Gu:2012ib}
Z.-C. Gu and X.-G. Wen, \emph{{Symmetry-protected topological orders for interacting fermions: Fermionic topological nonlinear {\ensuremath{\sigma}} models and a special group supercohomology theory}}, \href{https://doi.org/10.1103/PhysRevB.90.115141}{\emph{Phys. Rev. B} {\bfseries 90} (2014) 115141} [\href{https://arxiv.org/abs/1201.2648}{{\ttfamily 1201.2648}}].

\bibitem{Kitaev2009}
A.~Kitaev, \emph{Periodic table for topological insulators and superconductors},  in \emph{AIP Conference Proceedings}, vol.~1134, pp.~22--30, 2009, \href{https://arxiv.org/abs/0901.2686}{{\ttfamily 0901.2686}}, \href{https://doi.org/10.1063/1.3149495}{DOI}.

\bibitem{Schnyder:2008tya}
A.~Schnyder, S.~Ryu, A.~Furusaki and A.~Ludwig, \emph{{Classification of topological insulators and superconductors in three spatial dimensions}}, \href{https://doi.org/10.1103/PhysRevB.78.195125}{\emph{Phys. Rev. B} {\bfseries 78} (2008) 195125} [\href{https://arxiv.org/abs/0803.2786}{{\ttfamily 0803.2786}}].

\bibitem{Kapustin:2014dxa}
A.~Kapustin, R.~Thorngren, A.~Turzillo and Z.~Wang, \emph{{Fermionic Symmetry Protected Topological Phases and Cobordisms}}, \href{https://doi.org/10.1007/JHEP12(2015)052}{\emph{JHEP} {\bfseries 12} (2015) 052} [\href{https://arxiv.org/abs/1406.7329}{{\ttfamily 1406.7329}}].

\bibitem{Gaiotto:2015zta}
D.~Gaiotto and A.~Kapustin, \emph{{Spin TQFTs and fermionic phases of matter}}, \href{https://doi.org/10.1142/S0217751X16450445}{\emph{Int. J. Mod. Phys. A} {\bfseries 31} (2016) 1645044} [\href{https://arxiv.org/abs/1505.05856}{{\ttfamily 1505.05856}}].

\bibitem{Ryu:2010zza}
S.~Ryu, A.~P. Schnyder, A.~Furusaki and A.~W.~W. Ludwig, \emph{Topological insulators and superconductors: ten-fold way and dimensional hierarchy}, \href{https://doi.org/10.1088/1367-2630/12/6/065010}{\emph{New J. Phys.} {\bfseries 12} (2010) 065010} [\href{https://arxiv.org/abs/0912.2157}{{\ttfamily 0912.2157}}].

\bibitem{Inamura:2021szw}
K.~Inamura, \emph{{On lattice models of gapped phases with fusion category symmetries}}, \href{https://doi.org/10.1007/JHEP03(2022)036}{\emph{JHEP} {\bfseries 03} (2022) 036} [\href{https://arxiv.org/abs/2110.12882}{{\ttfamily 2110.12882}}].

\bibitem{Inamura:2021wuo}
K.~Inamura, \emph{{Topological field theories and symmetry protected topological phases with fusion category symmetries}}, \href{https://doi.org/10.1007/JHEP05(2021)204}{\emph{JHEP} {\bfseries 05} (2021) 204} [\href{https://arxiv.org/abs/2103.15588}{{\ttfamily 2103.15588}}].

\bibitem{Inamura:2024jke}
K.~Inamura and S.~Ohyama, \emph{{1+1d SPT phases with fusion category symmetry: interface modes and non-abelian Thouless pump}},  \href{https://arxiv.org/abs/2408.15960}{{\ttfamily 2408.15960}}.

\bibitem{Li:2024fhy}
Y.~Li and M.~Litvinov, \emph{{Non-invertible SPT, gauging and symmetry fractionalization}},  \href{https://arxiv.org/abs/2405.15951}{{\ttfamily 2405.15951}}.

\bibitem{Maeda:2025rxc}
J.~Maeda and T.~Oishi, \emph{{N-ality symmetry and SPT phases in (1+1)d}}, \href{https://doi.org/10.1007/JHEP12(2025)063}{\emph{JHEP} {\bfseries 12} (2025) 063} [\href{https://arxiv.org/abs/2504.20151}{{\ttfamily 2504.20151}}].

\bibitem{Lu:2025yru}
D.-C. Lu and Z.~Sun, \emph{{Intrinsic NISPT Phases, igNISPT Phases, and Mixed Anomalies of Non-Invertible Symmetries}},  \href{https://arxiv.org/abs/2511.01965}{{\ttfamily 2511.01965}}.

\bibitem{Cao:2025qhg}
W.~Cao, M.~Yamazaki and L.~Li, \emph{{Duality Viewpoint of Noninvertible Symmetry-Protected Topological Phases}}, \href{https://doi.org/10.1103/4zfz-x9xh}{\emph{Phys. Rev. Lett.} {\bfseries 136} (2026) 040402} [\href{https://arxiv.org/abs/2502.20435}{{\ttfamily 2502.20435}}].

\bibitem{Lu:2025rwd}
D.-C. Lu, F.~Xu and Y.-Z. You, \emph{{Strange correlator and string order parameter for non-invertible symmetry protected topological phases in 1+1d}},  \href{https://arxiv.org/abs/2505.00673}{{\ttfamily 2505.00673}}.

\bibitem{Aksoy:2025rmg}
{\"O}.~M. Aksoy and X.-G. Wen, \emph{{Phases with non-invertible symmetries in 1+1D - symmetry protected topological orders as duality automorphisms}},  \href{https://arxiv.org/abs/2503.21764}{{\ttfamily 2503.21764}}.

\bibitem{Warman:2024lir}
A.~Warman, F.~Yang, A.~Tiwari, H.~Pichler and S.~Schafer-Nameki, \emph{{Categorical Symmetries in Spin Models with Atom Arrays}}, \href{https://doi.org/10.1103/249m-m8wq}{\emph{Phys. Rev. Lett.} {\bfseries 135} (2025) 206503} [\href{https://arxiv.org/abs/2412.15024}{{\ttfamily 2412.15024}}].

\bibitem{Meng:2024nxx}
C.~Meng, X.~Yang, T.~Lan and Z.~Gu, \emph{{Non-invertible SPTs: an on-site realization of (1+1)d anomaly-free fusion category symmetry}},  \href{https://arxiv.org/abs/2412.20546}{{\ttfamily 2412.20546}}.

\bibitem{Tantivasadakarn:2017xbg}
N.~Tantivasadakarn, \emph{{Dimensional Reduction and Topological Invariants of Symmetry-Protected Topological Phases}}, \href{https://doi.org/10.1103/PhysRevB.96.195101}{\emph{Phys. Rev. B} {\bfseries 96} (2017) 195101} [\href{https://arxiv.org/abs/1706.09769}{{\ttfamily 1706.09769}}].

\end{thebibliography}\endgroup

\end{document}